\documentclass[]{aastex631}

\newcommand{\kms}{km\,s$^{-1}$}
\newcommand{\change}[1]{#1}

\graphicspath{{./}{./}}

\begin{document}


\title{TSD: An inverse problem approach for recovering the exoplanetary atmosphere transmission spectrum from high-resolution spectroscopy}

\author[0000-0001-5742-7767]{Nikolai Piskunov}
\affiliation{Astronomy Division, Dept. of Physics and Astronomy, 
Uppsala University, Box 516, 75120 Uppsala, Sweden}

\author[0000-0003-3486-853X]{Adam D. Rains}
\affiliation{Astronomy Division, Dept. of Physics and Astronomy, 
Uppsala University, Box 516, 75120 Uppsala, Sweden}
\affiliation{Instituto de Astrofísica, Pontificia Universidad Católica de Chile, Av. Vicuña Mackenna 4860, 782-0436 Macul, Santiago, Chile}

\author[0000-0003-1739-3827]{Linn Boldt-Christmas}
\affiliation{Astronomy Division, Dept. of Physics and Astronomy,
Uppsala University, Box 516, 75120 Uppsala, Sweden}




\begin{abstract}

Our ability to observe, detect, and characterize exoplanetary atmospheres has grown by leaps and bounds over the last 20 years, aided largely by developments in astronomical instrumentation; improvements in data analysis techniques; and an increase in the sophistication and availability of spectroscopic models. Over this time, detections have been made for a number of important molecular species across a range of wavelengths and spectral resolutions. Ground-based observations at high resolution are particularly valuable due to the high contrast achievable between the stellar spectral continuum and the cores of resolved exoplanet absorption features. However, recovering such features without relying on a theoretical or empirical model of the exoplanet atmosphere remains a major hurdle in data analysis, with traditional methods being limited by both the choice of algorithm used to remove the non-exoplanetary components of the signal, as well as the accuracy of model template spectra used for cross-correlation. Here we present a new algorithm TSD (Transmission Spectroscopy Decomposition) formulated as an inverse problem that reconstructs the planetary transmission spectrum without recourse to an exoplanet-atmosphere template, minimizing the number of assumed and theoretically modelled components. Instead of cross-correlation with pre-computed template exoplanet spectra, we rely on high spectral resolution and instrument stability to distinguish between the stellar, exoplanetary, and telluric components and velocity frames in the sequence of absorption spectra taken during multiple transits. We demonstrate the performance of our new method using both simulated and real $K$ band observations from ESO’s VLT/CRIRES+ instrument, and present results obtained from two transits of the highly-inflated super-Neptune WASP-107 b which orbits a nearby K7V star. 
\end{abstract}



\section{Introduction}\label{sec:introduction}
Since the discovery of the first exoplanet orbiting a main sequence star, 51~Pegasi b, in 1995 \citep{1995Natur.378..355M}, astronomers have found over 6\,000 exoplanets\footnote{See NASA Exoplanet Archive: \url{exoplanetarchive.ipac.caltech.edu}.} around stars of many types. Transits, first detected by \citet{2000ApJ...529L..45C}, have proved the most efficient detection technique and account for some ${\sim}70$\% of them; combined with radial velocity (RV) measurements they also give the size and mass of the planet, and hence the bulk density that is the starting point for classification. On that basis whole classes absent from the Solar System have emerged, such as super-Earths, with terrestrial densities but $\sim$3--10 times the mass of the Earth, and mini-Neptunes, with thick hydrogen--helium atmospheres and radii of only $\sim$2--4 Earth radii. 51~Pegasi b itself exemplified the first such class, the hot Jupiters, absent from the Solar System but common around solar-type stars.

Due to their large radii, short orbital periods, and high temperatures, hot Jupiters have been the focus of many space- and ground-based attempts to derive the physical properties of their atmospheres such as chemical composition, vertical temperature and pressure profiles \citep[e.g.][]{sing_hubble_2008,birkby_detection_2013,sedaghati_detection_2017,boucher_characterizing_2021,prinoth_titanium_2022,brogi_roasting_2023,tsai_photochemically_2023,smith_combined_2024}. These studies have broadly proceeded along two lines: time-series observations---often space-based---at high-photometric precision (but low-spectral resolution), or ground-based observations at high-spectral resolution from highly-stable ground-based instruments. For transmission\footnote{These objects are also highly-amenable to \textit{emission} (i.e. dayside) observations, but we restrict ourselves to considering transmission observations only for the remainder of our work here.} observations, both methods take advantage of the fact that during transit, part of the stellar flux is blocked by the planet including the outer semitransparent layers that compose its atmosphere. This transparency is a function of its chemical composition, thickness, temperature, pressure, and wavelength, and obtaining time-series data over a range of wavelengths simultaneously thus may be sensitive to the properties of the planetary atmosphere. 

Photometric or low-resolution observations compare transit light curves in spectral windows chosen to match strong features of particular species, such as band heads of H$_2$O or CH$_4$. At these resolutions individual planetary lines are poorly resolved, which reduces the contrast between transparent and strongly absorbing wavelengths but improves the overall signal-to-noise ratio (S/N). The approach requires high-precision instruments and relies on the photometric stability of space-based telescopes, many planets having been characterized with Hubble (e.g. \citealt{charbonneau_detection_2003}, \citealt{Vidal-Madjar_detection_2004}, \citealt{Pont_detection_2008}, \citealt{Deming_infrared_2013}, \citealt{Kreidberg_detection_2015}, \citealt{Wakeford_neptune_2017}), Spitzer (e.g. \citealt{Tinetti_water_2007}, \citealt{desert_search_2009}, \citealt{Knutson_spitzer_2011}, \citealt{cowan_thermal_2012}), and more recently JWST (e.g. \citealt{2023Natur.614..664A}, \citealt{Rustamkulov_early_jwst}, \citealt{dyrek_wasp107_2024}, \citealt{2025MNRAS.540.2535A}). Although this method of characterization can be done with ground-based facilities (e.g. \citealt{bean_ground-based_2011}, \citealt{sing_gran_2011}, \citealt{2016A&A...587A..67L}) the achievable photometric stability---and with it the precision of the transit depth measured in each spectral bin---remains considerably better from space, where observations are free from the variability and telluric contamination of Earth's atmosphere. That precision, rather than spectral resolution, is what is expected to bring the atmospheres of Earth analogues within reach, while shorter-period planets can be targeted from the ground. Interpretation, however, requires a grid of planetary atmosphere models to match against the observations. 

One alternative method takes advantage of the high contrast provided by strong \textit{resolved} atomic or molecular absorption lines to distinguish the star and planet via the now-resolved Doppler shifts induced by the planet's orbital motion. This setup requires high spectral resolution, which at present is mostly available on ground-based telescopes. Both the optical and the near-infrared are used, and the choice between them follows from the atmosphere being studied: high-resolution optical spectroscopy has been very successful for ultra-hot giant planets, where the accessible features are lines of atoms and ions, whereas for cooler planets such as WASP-107 b the diagnostic features are molecular bands and the near-infrared is considerably more advantageous. It allows the use of the largest photon collecting power but suffers from contamination by telluric features in the Earth's atmosphere.\footnote{This can complicate and cause confusion during data analysis due to the fact that some of the strongest telluric species like H$_2$O, CO, and CO$_2$ are \textit{also} key target species in exoplanet atmospheres.} The high-resolution approach was pioneered by \citet{snellen_orbital_2010} and this method with some enhancements is actively used today for the analysis of the transmission spectroscopy data \citep[see review by][]{snellen_review_2025}. The core idea is that an observed transmission spectrum is composed of three main components---stellar, telluric, and exoplanetary---with the stellar and telluric signals being essentially static\footnote{In practice, the stellar lines experience a small (of order ${\sim10}$ m\,s$^{-1}$) RV shift due to changes in the barycentric velocity over the transit.} in wavelength space while the exoplanet transmission spectrum shifts with the orbital motion of the planet. Assuming high enough spectral resolution\footnote{For a typical hot Jupiter on a 5 day orbit at 0.05 AU distance from a host K-dwarf star the line-of-sight velocity changes during transit by as much as 12 \kms. Thus, the resolution should be significantly higher with corresponding requirements on instrument stability.} one can attempt to remove the stellar and telluric components but not the exoplanetary signal based on the assumption that stellar and telluric spectra only change in time but not in wavelength. The removal is typically done with an iterative detrending process. This is sometimes Principal Component Analysis (PCA), but most commonly the SYSREM algorithm is used (e.g. \citealt{tamuz_correcting_2005}, \citealt{birkby_detection_2013}), which is related to PCA but not equivalent to it: it removes systematic trends (stellar and telluric features) iteratively while weighting each measurement by its uncertainty. The detrended spectra---or residuals---contain noise and the planetary signal following its orbital Doppler shift (therefore, it is important to have a reliable and accurate orbital solution from RV observations). Detection of chemical species in the planetary atmosphere is then achieved by performing cross-correlation with a synthetic transmission template spectrum and comparing the trace of cross-correlation peaks throughout the transit with the prediction of the orbital solution.

The two methods are complementary. Flux-calibrated spectrophotometry covers a large spectral range, often at wavelengths inaccessible from the ground, giving simultaneous information on many molecular species. Critically, it also preserves the continuum of the planetary transmission spectrum---typically destroyed by the detrending used at high resolution---which encodes the temperature--pressure and cloud structure of the atmosphere. High-resolution spectroscopy can instead distinguish molecular and \textit{atomic} species whose lines overlap in wavelength, given an adequate template for the cross-correlation step, enabling relative abundance measurements.\footnote{For a good summary of these advantages---and challenges---see the introduction of \citealt{Brogi_retrieving_2019}.} By resolving individual lines, their shifts and their shapes, it can also probe the ``weather'' on these planets, such as extreme winds and day--night differences \citep[e.g.][]{snellen_orbital_2010,brogi_rotation_2016,seidel_into_2021,gandhi_retrieval_2023,cont_exploring_2024,seidel_vertical_2025,nortmann_crires_2025}. Both share one limitation: they use models already at the signal detection stage, where the adequacy of the models and their coverage of parameter space are hard to assess, which raises questions about the resulting parameter uncertainties (see e.g. \citealt{savel_uncertainty_2025} and discussion within).

Returning to the high-resolution technique, detrending followed by cross-correlation has limitations of its own. They include the loss of the exoplanet continuum, which constrains the temperature--pressure profile (e.g. \citealt{de_kok_detection_2013}) and clouds where present; a sensitivity to the accuracy and completeness of the line lists used to generate templates (e.g. \citealt{Hoeijmakers_search_2015}, \citealt{brogi_framework_2017}, \citealt{Brogi_retrieving_2019}, \citealt{Gandhi_molecular_2020}); the inability to combine transits until after cross-correlation, that is \textit{after} models have entered the analysis, chiefly because of the differing barycentric velocities but also inter-night instrumental and weather variations; the fact that detrending methods do not naturally `converge' and thus require some sort of statistical metrics, simulations, or injection tests to determine the optimal number of iterations or components to remove\footnote{In practice the number of principal components to remove depends on wavelength (e.g. the location and strength of stellar, telluric, and exoplanet absorption features), the weather (e.g. humidity affecting the strength or variability of H$_2$O lines), the timing of the transit itself (e.g. barycentric velocity and how close the telluric and stellar RV frames are), instrumental effects (e.g. time varying PSF or blaze), and the combination of star, planet, and species of interest (e.g. whether the star/tellurics/planet have species in common).} (e.g. \citealt{Cheverall_robustness_2023}, \citealt{cathal_2024_methods}); the influence on the resulting detection of apparently unrelated choices in observation and reduction, such as cadence (e.g. \citealt{2024A&A...683A.244B}), spectral normalization or pixel weighting; and, perhaps most importantly, that detrending destroys part of the planetary signal itself (e.g. \citealt{birkby_discovery_2017}, \citealt{meech_2022_gp}). This approach has been successfully applied to many planets for the better part of the last 15 years. These limitations motivate the exploration of complementary analysis techniques such as forward or inverse models (e.g. \citealt{aronson_using_2015}, \citealt{blain_formally_2024}), or Doppler tomography (e.g. \citealt{Watson_doppler_2019}, \citealt{Esparza-Borges_retrieving_2022}) which approach the problem from a different angle.

In this paper, we describe a new inverse method capable of extracting planetary transmission spectra from high-resolution transmission spectroscopy without the use of detrending algorithms or the reliance on theoretical exoplanet models in the signal detection phase. We build upon the inverse modelling work by \citet{aronson_using_2015} to develop a functional inverse problem framework and software code, capable of reliably extracting planetary transmission spectrum for warm and cool transiting exoplanets as observed with ground-based high-resolution spectrographs. In our numerical experiments, we consider $K$ band observations of our test system, the warm Neptune WASP-107 b, at realistic S/N as achievable by current generation instrumentation. We call the new method ``\textbf{T}ransmission \textbf{S}pectroscopy \textbf{D}ecomposition" or TSD.\footnote{Internally, TSD is also occasionally referred to as ``TueSDay".} TSD takes advantage of high spectral resolution, wavelength stability, and multiple transits to distinguish between the stellar, telluric, and planetary components---independently of any theoretical or empirical template of the exoplanet transmission spectrum. TSD is thus independent of the planetary model, though not of every model: it still assumes a stellar-atmosphere model for the limb-darkening (specific-intensity) relation, a parameterized telluric optical depth, and \textit{a priori} stellar, planetary, and barycentric velocities (see Section~\ref{sec:forward_model}).

The idea of separating superposed spectra based on their
Doppler shifts across a sequence of observations is of course not new: it is routinely
used for spectral disentangling of binary systems and Doppler tomography of transiting planets
\citep[e.g.][]{Watson_doppler_2019,Esparza-Borges_retrieving_2022}. The problem solved
here differs from those in four respects that are qualitative rather than incremental.
The separation involves three components rather than two, one of which, the telluric
absorption, enters multiplicatively through an exponential of the airmass. That component
is not static: its strength changes during and between transits and has to be solved for
rather than assumed constant. The contrast is far less favourable, the planetary features
that carry the atmospheric information 3 to 4 orders of magnitude weaker than the stellar
flux. Spectra contain dense forests of stellar and telluric lines whose mutual overlap differs
from transit to transit as the Doppler shifts change. That changing overlap is making impossible
reliable continuum normalization.

The paper is structured as follows: Section \ref{sec:forward_model} introduces our new inverse method, Section \ref{sec:simulator} gives an overview of our simulated transmission spectra formalism which we use to demonstrate our model's performance using simulated $K$ band transmission spectra matching ESO's VLT/CRIRES+ instrument; Section \ref{sec:spectra} describes our observational data and the results of testing our model on two real transits of WASP-107 b; and Section \ref{sec:discussion} contains our analysis of results, a discussion putting TSD in context with more traditional detrending methodologies, and conclusions to the paper.

\section{Forward model}\label{sec:forward_model}

In this Section, we will introduce the model $M_\lambda^j$ of an observed high-resolution spectrum of a transiting system taken in orbital phase number $j$, where $\lambda$ is the wavelength in the restframe of the spectrometer. The model involves stellar flux ($F_\lambda^\star$) diminished by a wavelength-dependent planet ``shadow'' $I\cdot P$, where $I$ is the specific intensity at the occulted position and $P$ the effective blocking area of the planet, including the semi-transparent layers of its atmosphere, in units of the stellar disk area. With $w_j$ carrying the phase dependence, $P$ is a function of wavelength alone. Stellar light modified by the planet is then absorbed by telluric features in Earth's atmosphere with optical depth $\tau_\lambda$ scaled by the airmass $z$. We also include scaling ($S_j$) and offset ($D_j$) terms accounting for the seeing conditions and data reduction respectively. Equation~\ref{eq: forward model} is written for one spectral segment, and the segments are solved independently, so $S$ and $D$ are determined per segment as well as per phase. This is what lets them absorb the order-to-order variation of the throughput and of the background level, which a single pair of numbers per exposure could not represent. Within a segment they are constant with wavelength: for the chromatic structure on that scale we rely on the stability and reproducibility of CRIRES$^+$ and of its reduction pipeline, the blaze being removed with the flat field, so that what remains to be absorbed is one level and one scale per segment. $M$, $F^\star$, $I$ and $D$ carry the units of the observed spectrum; $P$, $\tau$, $z$, $S$ and the Doppler factors below are dimensionless.
\begin{equation}
M_\lambda^j = \left[F_{\lambda\cdot(1-\gamma_j)}^\star - I_{\lambda\cdot(1-\beta_j)}^j\cdot P_{\lambda\cdot(1-\delta_j)}\cdot w_j\right]\cdot
e^{-\tau_\lambda\cdot z_j}\cdot S_j + D_j.
\label{eq: forward model}
\end{equation}

\noindent
The product $I\cdot P$ is scaled by the geometric weight $w_j$: zero outside transit, unity while the planetary disk lies entirely within the stellar disk, and equal to the fraction of the planetary disk overlapping the stellar disk during ingress and egress. It is precomputed for each phase from the orbital solution and the transit geometry and is not a fitted parameter. The same calculation returns the position of the centre of the blocked area, which during the partial phases does not coincide with the projected centre of the planet; it is that position which defines the limb distance $\mu_j$ at which the specific intensity is evaluated. In all expressions that follow, $w_j$ is carried inside the specific-intensity term.\\

\noindent
$\gamma$, $\beta$ and $\delta$ are unitless Doppler factors defining the shift in wavelength of the stellar flux, the specific intensity originating from the area blocked by the planet, and the planet transmission from their corresponding restframes to the observer reference frame. We assume the velocity to be positive when directed away from the observer (red shift) and therefore:
\begin{itemize}
\item $\gamma_j = (v_\mathrm{bary}+v_\mathrm{star})/c$ is a barycentric + radial velocity Doppler shift factor for the star ($c$ is the speed of light).
\item $\beta_j = (v_\mathrm{bary}+v_\mathrm{star}+v\sin i\cdot x_\mathrm{planet})/c$ differs from stellar velocity if the data resolves stellar rotation. $x_\mathrm{planet}$ is the distance between the projected planet position and the projection of the stellar rotation axis in units of the stellar radius. $x_\mathrm{planet}$ is negative for the hemisphere that moves toward the observer. $v\sin i$ is the line-of-sight component of the equatorial stellar rotation. The underlying assumption of solid rotation can be (if necessary) generalized to include differential rotation.
\item $\delta_j = (v_\mathrm{bary}+v_\mathrm{star}+v_\mathrm{planet}^j)/c$ is the planet's Doppler shift. In the simplest case of a circular orbit, the planetary component will be the line-of-sight component of the orbital velocity multiplied by $\sin\theta_j$, where $\theta_j$ is the orbital phase angle of exposure $j$ measured from mid-transit. We reserve $j$ for the exposure index throughout and use $\theta$ for the orbital phase.
\end{itemize}
With this sign convention, a spectral feature located at wavelength $\lambda_0$ in the stellar restframe is observed at $\lambda_0\cdot(1+\gamma_j)$, which is the notation used in Equation~\ref{eq: forward model} and throughout this paper.

In the rest of the paper the "planet shadow" $P$ is referred to as the ``effective blocking area". It is given by the ratio of radii squared $\left(R_\mathrm{planet}/R_\star\right)^2$ where $R_\mathrm{planet}$ is the effective radius of the planet blocking all radiation at a given wavelength. For the case of an Earth-like planet, it can be understood as the radius of the planet surface plus the height of the atmosphere that is optically thick for the grazing rays at a given wavelength. For gas planets, no such distinction exists, but the value of $P \geq 0$ in both cases. Atmospheric transmission is, of course, a function of wavelength and therefore the subject of Doppler shifts due to the orbital motion of the planet. For all the orbital phases where planetary footprint is fully inside the stellar disk, this Doppler shift is the only dependence on the orbital phase (assuming a spherically symmetric atmosphere). $P_{\lambda\cdot(1-\delta_j)}$ changes dramatically as the planet crosses the stellar limb, and goes to zero outside the transit.

The exponential multiplier to the square bracket is the description of the Earth's telluric absorption. Here, $\tau_\lambda$ is the monochromatic optical path through the Earth atmosphere in the zenith direction (airmass unity) and $z_j$ is the airmass of the target at every observed phase. The choice of optical path naturally constrains the telluric features to absorption, preventing instabilities in the inverse solver (see Section~\ref{sec:forward_model:minimisation}). Finally, the phase-dependent factor $S_j$ and offset $D_j$ are there to correct for variable observing conditions, mostly affected by the airmass, seeing, wind, slit losses and data reduction. Ideally, these vectors should be smooth function of the airmass/time but in real life this is seldom the case.

When writing Equation~\ref{eq: forward model}, we make several simplifying assumptions. We assume that the stellar flux $F$ and the planet blocking area $P$ are constant in time (i.e. no phase dependence), and the scaling factor $S$ and offset $D$ change with phase but are the same for all wavelengths within each spectral segment. For different transits (i.e. nights), telluric absorption may vary, but for a single transit it does not change in time except due to the variable airmass $z_j$. This is a strong approximation as e.g. the water content can change during a night independently of the airmass braking our assumption. For the two transits that we choose to illustrate TSD performance we have tested the variation of $\tau_\lambda$ directly, and it exceeds the noise of the data in only two phases, both on the second night and both outside the transit, where the geometric weight $w_j$ is zero and the blocking area $P$ cannot be affected. The residuals in Figure~\ref{fig:PSF stability} clearly show these two phases. Relaxing the assumption, with water treated as a component separate from other species, is among the developments listed in Section~\ref{sec:discussion}. Finally, we assume that the radial components of all velocities (star, planet, barycentric) are known for each phase \textit{a priori}. These assumptions break down in the case of stellar activity or spots, exoplanet asymmetries or time and velocity resolved ``weather'', or large Earth-humidity variations within a single night that would cause H$_2$O telluric features to be time variable. We discuss the effect of these limitations further in Section \ref{sec:discussion:limitations}.

One further approximation concerns the instrumental profile. Equation~\ref{eq: forward model} is written at the resolution of the instrument: the unknown functions $F^\star$, $P$ and $\tau$ are reconstructed on the pixel grid of the observations and are combined by multiplication, so that no convolution operator appears in the model or in the update equations. Strictly, the observed spectrum is the convolution of the \textit{product} of the three components with the line-spread function, and the convolution does not commute with the multiplication. The two differ only where the components have structure on the scale of the resolution element. Because we do not deconvolve the observations, the recovered functions should be understood as the component spectra \textit{at the resolution of the instrument}. Since the components move around due to Doppler shift on sub-resolution scales this may produce errors. The magnitude of these errors was examined: in the simulations described in Section~\ref{sec:simulator:methodology} the instrumental profile is applied to the product of the components, exactly as the real instrument does, while the inversion multiplies component spectra already at instrumental resolution, so the reconstructions presented in Section~\ref{sec:simulator:performance} include this error in full. An explicit treatment, in which the components are reconstructed at higher resolution and convolved before comparison with the data, may be required for spectrographs whose line-spread function varies appreciably with time. For CRIRES+ the instrumental profile is stable and well sampled by 3.5 pixels, as we demonstrate for our observations in Section~\ref{sec:spectra} (Figure~\ref{fig:PSF stability}).

With these assumptions, we can now list the unknown functions: stellar flux $F$, planet blocking area $P$, intensity scaling factor $S$, background offset $D$, and optical thickness of telluric absorption $\tau$ for every transit. The inclusion of multiple transits is not just desirable. It is crucial for success: the changes of Doppler shift during a single transit are not sufficient to distinguish between the telluric and the stellar lines, thus creating a degeneracy. Change in barycentric correction \textit{between} transits can shift stellar features sufficiently to remove this ambiguity.

\subsection{Minimisation problem and unknown functions}\label{sec:forward_model:minimisation}

The discrepancy function $\Phi$ measures the differences between the model $M_\lambda^j$ and the observations $O_\lambda^j$:
\begin{equation}
\Phi \equiv \sum_{\lambda,j} \omega_\lambda^j \left[M_\lambda^j - O_\lambda^j\right]^2.
\label{eq: minimisation problem}
\end{equation}

\noindent
where $\omega_\lambda^j$ are optional inverse-variance weights for the individual data points. The objective function actually minimized by the current implementation of TSD is Equation~\ref{eq: minimisation problem} with uniform weights ($\omega_\lambda^j=1$), augmented by the Tikhonov regularization terms \citep{Tikhonov1995} for $F^\star$ and $P$, specified in
Section~\ref{sec:forward_model:degeneracies}; the update equations derived in Appendix~\ref{app:linearization} and summarized in section~\ref{sec:forward_model:solving} therefore carry no explicit $\omega$. We emphasize that the strong attenuation in the cores of telluric lines is not a substitute for inverse-variance weighting: it acts through the forward model itself so that heavily absorbed wavelengths carry little leverage in the stellar flux and blocking area updates irrespective of the choice of $\omega$. The per-pixel measurement uncertainties delivered by the reduction pipeline are not used to weight the solution; they enter only the calculation of the reduced $\chi^2$ and the Monte-Carlo error-propagation exercise of Section~\ref{sec:spectra}. A full inverse-variance weighting, together with a likelihood-based treatment of correlated and wavelength-dependent noise, is a natural extension of the present formulation; we return to this point when discussing the reduced $\chi^2$ values in Section~\ref{sec:discussion}. Now, we will attempt to reconstruct the unknown functions that minimize $\Phi$.

To inform the sequence of our solving procedure described later in Section \ref{sec:forward_model:solving}, it is useful to take a closer look at each of the unknown functions $F$, $P$, $T$, $S$, and $D$ before proceeding. The 1D spectral vectors $F$, $P$, and $\tau$ have dimensions of the wavelength grid ($N_\lambda$), and the 1D time-series vectors $S$ and $D$ have dimensions equivalent to the number of phases ($N_\mathrm{phase}$). They are to be reconstructed from a 2D array of observations that has $N_\lambda\times N_\mathrm{phase}$ elements and, thus, the problem described by Equation~\ref{eq: minimisation problem} is not under-determined. This does not by itself guarantee a unique solution: the model is non-linear and contains near-degenerate directions, which we analyse in Section~\ref{sec:forward_model:degeneracies}, where we also define what we mean by uniqueness in this context. Moreover, elements of unknown vectors are only affected by a subset of observational data reducing cross-talk and improving the chances of obtaining a unique solution: as shown in Section~\ref{sec:simulator:performance}, the reconstruction does approach the input spectrum as the S/N and the number of transits increase. For example, any given element of the telluric spectrum---which does not shift with barycentric velocity---has a one-to-one correspondence with the wavelength elements of the observations meaning that each telluric absorption element is strictly defined by the combined $N_\mathrm{phase}$ data points. The stellar flux vector, which shifts with barycentric velocity, is only affected by a couple of wavelength elements per observed transit. An exoplanet atmosphere transmission element, which sees velocity shifts from both the barycentric and exoplanet motion, is affected by 10--20 pixels in each transit and these shift following the stellar lines between transits.

The telluric optical depth is not measured outside the transit and interpolated through it. It is one of the unknown functions, solved for jointly with the others and scaled to each phase by that phase's airmass, subject to the approximation discussed above; spectra taken before and after the transit serve only to provide an initial guess and to check the result. For multiple transits, however, we cannot assume this function to be the same and so we must determine the telluric absorption vector for each transit separately. Similarly, spectra taken before and after the transit and corrected for telluric absorption can provide a good initial guess for the stellar flux spectrum $F_\lambda$.
For the planetary blocking area $P$, we know that it is both positive and a function of wavelength, so a suitable initial guess could then be a constant value set to the squared ratio of the photometric planet and stellar radii --- the geometric area ratio, the wavelength-independent level about which $P_\lambda$ varies. Within a spectral segment the scaling $S_j$ and offset $D_j$ vectors are functions of the orbital phase only, and in this sense they are ``orthogonal'' to the flux, tellurics, and planetary blocking area, which are functions of wavelength.

\subsection{Specific intensity}\label{sec:forward_model:specific intensity}

The remaining unknown function is the specific intensity. To obtain $I_\lambda^j$ we can rely on a model atmosphere and spectral synthesis for the star. Orbital phase $j$ defines the limb distance parameter $\mu$ on the stellar disk, at which the specific intensity should be computed, where $\mu$ is the cosine of the angle between the local normal and the direction to the observer at the eclipsed part of the stellar surface. Thus, our procedure consists of the following steps:
 \begin{enumerate}
\item Find fundamental parameters of the host star, e.g. using SME (Spectroscopy Made Easy, \citealt{SME2}) with high-quality spectra and any other available non-spectroscopic constraints (asteroseismology, parallax, accurate photometry, etc.).
\item Compute a model atmosphere with the derived parameters.
\item Compute the disk-integrated flux spectrum.
\item Verify that it reproduces the observed spectrum, both in the range used to determine the parameters and, separately, in the wavelength range of the transit observations, which need not have entered the fit.
\item Compute $\mu$ positions for all phases (i.e. exposures) during the transit where the planet's shadow is fully or partially inside the stellar disk.
\item Compute synthetic specific intensities for the $\mu$ angles derived above.
\end{enumerate}
The resulting intensities will carry all the problems of spectral synthesis.

Alternatively, a specific intensity can be derived from the flux using a limb-darkening law \citep[e.g.][ and the references within]{2024NatAs...8..929K}. The theory of stellar atmospheres predicts the monochromatic intensity to be a smooth function of the limb distance $\mu$, frequently approximated with the following expression
\citep{2000A&A...363.1081C}:
\begin{equation}
I(\mu)/I(1) = 1-\sum_{k=1}^K a_k\cdot (1-\mu^{k/2})=1-\sum_{k=1}^K a_k+\sum_{k=1}^K a_k\cdot \mu^{k/2},
\label{eq: limb darkening}
\end{equation}
where $a_k$ are the fitting coefficients. Typically truncating summation at $K$ equal to 4 is considered sufficient to achieve better than 0.1\,\%\ precision \citep{2015A&A...573A..90M}. Assuming a non-rotating star without any active regions, we can also compute the monochromatic flux by disk integrating the specific intensity:
\begin{equation}
F=2\pi\int_0^1 I(\mu) \mu d\mu = \pi\cdot I(1)\cdot\left(1-\sum_{k=1}^K \frac{k\cdot a_k}{k+4}\right),
\label{eq: flux and specific intensity}
\end{equation}
which gives us the relation between the flux and the central intensity. Now we can use the flux in
Equation \ref{eq: limb darkening} instead of $I(1)$:
\begin{equation}
I(\mu) = \left. F\cdot\left(1-\sum_{k=1}^K a_k+\sum_{k=1}^K a_k\cdot \mu^{k/2}\right)\middle/
\pi\cdot\left(1-\sum_{k=1}^K \frac{k\cdot a_k}{k+4}\right).\right.
\label{eq: specific intensity}
\end{equation}

\begin{figure}[hbt!]
    \centering
    \includegraphics[width=0.95\linewidth, trim=2cm 5cm 2cm 11cm]{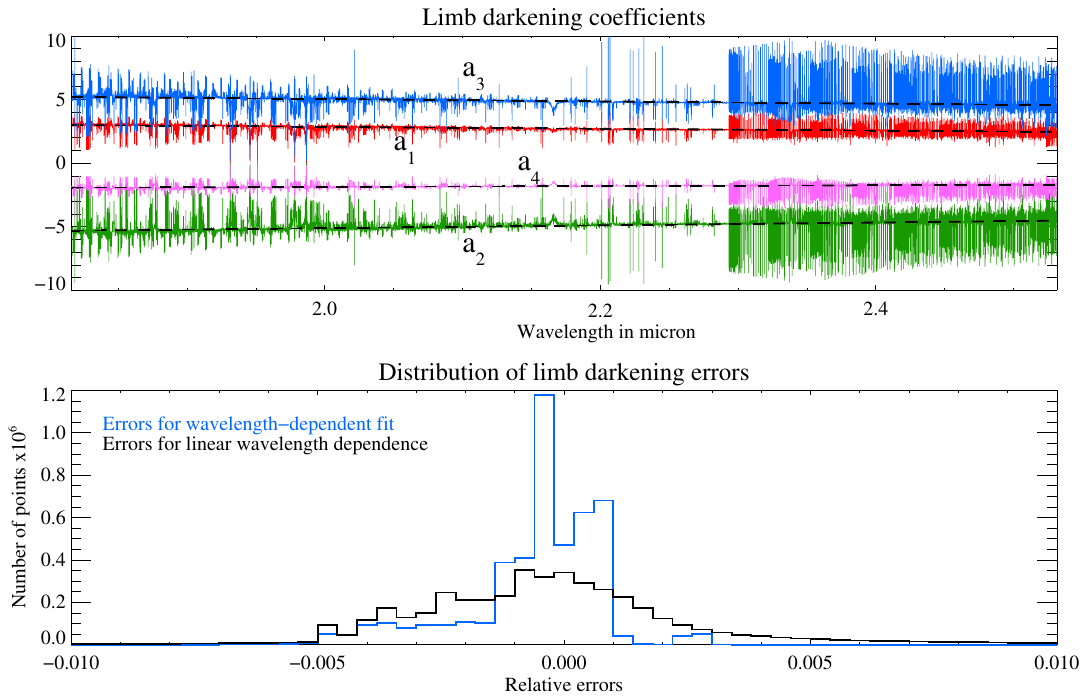}
    \caption{\textbf{Top}: Monochromatic limb darkening coefficients as in Equation~\ref{eq: limb darkening} computed for specific intensities in the near-infrared using a MARCS model atmosphere \citep{gustafsson_grid_2008}. The parameters of the model are the following: $T_\mathrm{eff}=3340\,$K, $\log g=5$, $\mathrm{[Fe/H]}=-0.34$, $v_\mathrm{macro}=1$\,\kms. Limb darkening was calculated for wavelength sampling at resolution R=100$\,$000 and for 49 limb distances ($\mu$ angles) with decreasing spacing toward the limb. Dashed black lines indicate the linear fit for each coefficient. \textbf{Bottom:} The distribution of relative errors when reconstructing specific intensities using wavelength-dependent limb darkening coefficients (in black) and their linear fit (in blue). In the first case all wavelength and disk points are reproduced to better than 1\,\%. For the blue histogram 94\,\%\ of points are reproduced to better than 1\,\%. These points are not random: they are located in the cores of the
 strongest stellar lines making TSD procedure less robust at these wavelengths.}
    \label{fig:limb darkening}
\end{figure}

Most of our planet hosts are slow rotators, meaning that rotational broadening is not resolved by our observations and we can safely assume zero rotation. Under this assumption, the $\gamma_j$ and $\beta_j$ Doppler factors are identical, reducing the number of unique velocity frames when modeling these targets. Slow rotation is also statistically associated with a low level of magnetic activity (see e.g. \citealt{dynamo_1996ApJ}, \citealt{rotationM_2016MNRAS}, \citealt{rotation_2018}, \citealt{rotation_2021}), but it does not guarantee the absence of active regions along the transit trajectory. Nor does it imply that the centre-to-limb variation of the line profiles can be ignored. That effect is retained in the model through the wavelength dependence of the limb-darkening coefficients: because the $a_k$ vary with wavelength, the line cores and the neighbouring continuum are limb-darkened by different amounts, so the occulted spectrum changes shape with $\mu_j$ and not merely in scale. The accuracy of that representation is the error budget quantified below.

Our aim here is to remove from the forward model the last unknown function that cannot be derived directly from observations. The advantage of using limb darkening as opposed to explicit calculation of specific intensity from model atmospheres is that the specific intensity is related to the observed flux, making it internally consistent. The downside is that the assumptions we make may hold for some systems but not for the others; however, slow rotation means that our observations do not resolve the rotational broadening of spectral lines. For example, for an instrument with resolving power $R$ = 100\,000, stellar rotation $v\sin i$ less than 1.5 km\,s$^{-1}$ will not change the measured line profiles. This is a limiting factor of the method as currently implemented, but we will properly account for stellar rotation in future versions of TSD.

The remaining fundamental problem is that the limb-darkening coefficients $a_k$ strongly depend on the opacity variation and thus on the wavelength. The coefficients could be determined empirically (i.e. as free parameters) or theoretically, but both methods have problems. If we were to empirically determine wavelength-dependent limb-darkening coefficients, we would risk ending up with a poorly constrained problem. On the other hand, model atmosphere-based coefficients are known not to perfectly match the reality of the Sun except when using 3D hydrodynamical models \citep{2015A&A...573A..90M}. This is particularly true for the blue part of the spectrum and close to the solar limb. Still, even without a 3D model this approach gives a very good fit to specific intensities for most of the stellar disk and allows us to assess the wavelength dependence of the coefficients for individual targets. An example based on the parameters of the planet-host star WASP-107 is shown in Figure~\ref{fig:limb darkening}. The selected spectral interval in the $K$ band is popular for high-resolution transmission spectroscopy of exoplanets because it contains spectral features of several interesting molecules, such as H$_2$O, CO, CO$_2$, NH$_3$, and CH$_4$. The derived coefficients are strongly variable on a small scale (as shown by the color lines on the top panel of Figure~\ref{fig:limb darkening}), but because the summation components in Equation~\ref{eq: specific intensity} are similar in magnitude and the odd and even coefficients have opposite signs, the wavelength dependence of the sum is largely reduced. In fact, accounting for the linear trend of the limb darkening coefficients already captures most of the effect. The distribution of relative errors between the specific intensities, computed from a MARCS atmospheric model \citep{gustafsson_grid_2008} and our approximation using $K=4$ shows that accounting for linear dependence brings the errors well below 1\,\%\ for the majority (94\,\%) of disk positions and wavelengths (bottom panel of Figure~\ref{fig:limb darkening}). This is only slightly worse than using the exact wavelength dependence of the limb darkening coefficients, and is comparable to the expected approximation accuracy of Equation~\ref{eq: specific intensity} with $K=4$. We note that larger deviations occur close to the limb where the planet spends only a small fraction of the transit.

TSD solves for the stellar flux and evaluates the specific intensity from it through Equation~\ref{eq: specific intensity}, with the four limb-darkening coefficients taken as linear functions of wavelength. Specific intensities are not interpolated from a model-atmosphere grid during the inversion; the model atmosphere enters only when we fit the coefficients and their wavelength trends.

The error this introduces can be quantified against the errors of the observations. The relative deviations between the synthetic specific intensities and the analytic representation with linear coefficients (blue histogram in the bottom panel of Figure~\ref{fig:limb darkening}) have a root-mean-square of $0.5\,\%$, to be compared with a mean relative uncertainty of $0.8\,\%$ per pixel in the observed spectra analysed here: the stellar-model term is smaller than the photon noise by a factor of $1.6$. Two further properties of this error matter for the way it propagates. First, it is not white: the deviations are correlated with the stellar spectral features, being largest in the cores of strong lines where the intensity varies most steeply with $\mu$, so they are not reduced by averaging over wavelength in the way the photon noise is. Second, the intensity enters the forward model only through the product $I_\lambda^j\cdot P_\lambda$, so to first order a relative error $\epsilon$ in the intensity produces a relative error of the same order in the recovered blocking area. With $\epsilon=0.005$ and a transit depth of about $2\,\%$ this amounts to roughly $1\times10^{-4}$ in units of the stellar disk area, more than an order of magnitude below the per-pixel uncertainty of the reconstruction derived in Section~\ref{sec:uncertainty}. We emphasise that this is a first-order estimate of the magnitude of the effect; its actual consequence for the recovered transmission spectrum is measured end-to-end by the simulations, which build the observations from synthetic specific intensities and let TSD reconstruct them with the analytic relation (Section~\ref{sec:simulator:methodology}).

Accepting linear wavelength dependence of the limb darkening coefficients, restricting the approximation to four terms, and introducing new notations for the nominator $L$ and denominator $N$ for the ratio in the RHS of Equation~\ref{eq: specific intensity}, we can re-write our model without using specific intensity explicitly:
\begin{eqnarray}
N_\lambda&=&\pi\cdot\left[1-\sum_{k=1}^4\frac{k\cdot a(\lambda)_k}{k+4}\right], \nonumber \\
L_{\lambda,j}&=&1-\sum_{k=1}^4 a(\lambda)_k\cdot(1-\mu_j^{k/2}), \\
M_\lambda^j &=& \left[F_{\lambda\cdot(1-\gamma_j)}^\star - F_{\lambda\cdot(1-\beta_j)}^\star\cdot
\frac{L_{\lambda,j}}{N_\lambda}\cdot P_{\lambda\cdot(1-\delta_j)}\right]\cdot e^{-\tau_\lambda\cdot z_j}\cdot S_j+D_j. \nonumber
\label{eq: final specific intensity}
\end{eqnarray}

We selected to not apply Doppler shift to the limb-darkening components: the slope of the wavelength dependence in our spectral region is so small that for typical values of $\gamma$ and $\beta$, the relative changes to $L$ and $N$ are below $10^{-7}$ and can be safely ignored.

\subsection{Geometry of the planetary shadow}\label{sec:forward_model:linearization}

Since the phase dependence of the shadow is carried by the geometric weight $w_j$, the blocking area that remains to be solved for is a function of wavelength alone. However, this is only partly correct as the atmosphere becomes a variable transparency circle due to several factors (day- and night-side differences, atmospheric refraction, etc.), but we note that the partial eclipse phases are less sensitive to $P$ and hence not crucial for recovery of the planet blocking area, making this an acceptable compromise.

A related approximation concerns the limb distance itself. Within one exposure the occulted region is represented by the single limb distance $\mu_j$ of the centre of the blocked area, rather than by integrating the specific intensity over that area. Two considerations bound the resulting error. The exposure times were chosen so that the planet advances by no more than $0.66$ of its own radius during an exposure while still reaching a signal-to-noise ratio above $110$, which limits how far the occulted region can smear within a single spectrum. More directly, we have compared the single-$\mu$ treatment with explicit integration of the specific intensity over the occulted area. For the inner part of the transit, within $0.9$ stellar radii of disk centre, the two agree to better than one part in $10^7$, which is five orders of magnitude below the uncertainty of the recovered transmission spectrum; all but three of our in-transit exposures fall in that region. The three exceptions lie close to ingress and egress, where the approximation is less accurate but the occulted area is also smallest, so those exposures carry correspondingly little weight in the solution.

The linearization of Equation~\ref{eq: forward model} and the derivation of the update equations for $S_j$, $D_j$, $F_\lambda^\star$, $P_\lambda$ and $\tau_\lambda$ are given in Appendix~\ref{app:linearization}; the results are collected in Section~\ref{sec:forward_model:solving}.


\subsection{Solving the equations}\label{sec:forward_model:solving}

Here are the equations we need to solve for the unknown function. The derivation can be found in Appendix~\ref{app:linearization}.
Forward model:
\begin{eqnarray}
N&=&\pi\cdot\left(1-\sum_{k=1}^4\frac{k\cdot a_k}{k+4}\right) \label{eq: summary norm} \\
L_j&=&1-\sum_{k=1}^4 a_k\cdot(1-\mu_j^{k/2}) \\
M_\lambda^j &=&\left[F_{\lambda\cdot(1-\gamma_j)}^\star - F_{\lambda\cdot(1-\beta_j)}^\star\frac{L_j}{N}\cdot P_{\lambda\cdot(1-\delta_j)}\right]\cdot
e^{-\tau_\lambda\cdot z_j}\cdot S_j + D_j.
\end{eqnarray}
Stellar flux:
\begin{equation}
F_\lambda^\star = \frac{\sum_j \left[O_{\lambda\cdot(1+\gamma_j)}^j-D_j\right] \cdot 
\left[1 - \frac{L_j}{N} \cdot P_{\lambda\cdot(1-\delta_j+\gamma_j)}\right]
\cdot e^{-\tau_{\lambda\cdot(1+\gamma_j)}\cdot z_j}\cdot S_j}
{\sum_j \left[1 - \frac{L_j}{N} \cdot 
P_{\lambda\cdot(1-\delta_j+\gamma_j)}\right]^2
\cdot e^{-2\tau_{\lambda\cdot(1+\gamma_j)}\cdot z_j}\cdot {S_j}^2}. \label{eq: summary flux}
\end{equation}

\noindent
Slit losses and background:
\begin{eqnarray}
S_j\cdot& \sum_\lambda &\left[F_{\lambda\cdot(1-\gamma_j)}^\star - F_{\lambda\cdot(1-\beta_j)}^\star\cdot\frac{L_j}{N}\cdot
P_{\lambda\cdot(1-\delta_j)}\right]^2\cdot e^{-2\tau_\lambda\cdot z_j} + \nonumber \\
+D_j\cdot& \sum_\lambda &\left[F_{\lambda\cdot(1-\gamma_j)}^\star -
F_{\lambda\cdot(1-\beta_j)}^\star\cdot\frac{L_j}{N}\cdot
P_{\lambda\cdot(1-\delta_j)}\right]\cdot e^{-\tau_\lambda\cdot z_j} = \label{eq: summary SD first} \\
=& \sum_\lambda& O_\lambda^j\left[F_{\lambda\cdot(1-\gamma_j)}^\star - F_{\lambda\cdot(1-\beta_j)}^\star\cdot\frac{L_j}{N}\cdot
P_{\lambda\cdot(1-\delta_j)}\right]\cdot e^{-\tau_\lambda\cdot z_j} \nonumber \\
S_j\cdot& \sum_\lambda &\left[F_{\lambda\cdot(1-\gamma_j)}^\star - F_{\lambda\cdot(1-\beta_j)}^\star\cdot\frac{L_j}{N}\cdot
P_{\lambda\cdot(1-\delta_j)}\right]\cdot e^{-\tau_\lambda\cdot z_j} + 
D_j\cdot\sum_\lambda 1 = \sum_\lambda O_\lambda^j. \label{eq: summary SD second}
\end{eqnarray}
Telluric optical depth:
\begin{equation}
\Delta \tau_\lambda = \frac{\sum_j \left(M_\lambda^j - O_\lambda^j\right)
\cdot \left(M_\lambda^j - D_j\right)\cdot z_j}
{\sum_j \left(2M_\lambda^j - O_\lambda^j - D_j\right)\cdot \left(M_\lambda^j - D_j\right)\cdot z_j^2}. \label{eq: summary tau}
\end{equation}
Planet blocking function:
\begin{equation}
P_\lambda = \frac{\sum_j
\left[F_{\lambda\cdot(1-\gamma_j+\delta_j)}^\star e^{-\tau_{\lambda\cdot(1+\delta_j)}\cdot z_j}
S_j - O_{\lambda\cdot(1+\delta_j)}^j + D_j\right]
F_{\lambda\cdot(1-\beta_j+\delta_j)}^\star\cdot \frac{L_j}{N}
\cdot e^{-\tau_{\lambda\cdot(1+\delta_j)}\cdot z_j} S_j}
{\sum_j\left[
F_{\lambda\cdot(1-\beta_j+\delta_j)}^\star\cdot \frac{L_j}{N}\right]^2\cdot 
e^{-2\tau_{\lambda\cdot(1+\delta_j)}\cdot z_j}\cdot {S_j}^2}. \label{eq: summary P}
\end{equation}
The unknowns are the optical thickness of telluric absorption $\tau_\lambda$, the stellar flux $F_\lambda^\star$, the planetary blocking area $P_\lambda$, and the scaling and offset parameters of observations $S_j$ and $D_j$.

The full system cannot be solved simultaneously, so we set up an iterative scheme starting with computing stellar flux (Equations~\ref{eq: summary norm}--\ref{eq: summary flux}), then correcting telluric optical depths (Equation~\ref{eq: summary tau}), updating scaling/offset (Equations~\ref{eq: summary SD first}--\ref{eq: summary SD second}), and, finally, deriving the planet blocking area (Equation~\ref{eq: summary P}). We use regularization in the wavelength dimension for $F_\lambda^\star$, $\tau_\lambda$, and $P_\lambda$ to avoid possible division by zero and high-frequency noise. In addition, the solvers constrain the telluric optical depth to $0 \leq \tau_\lambda \leq 20$ and the blocking area to the range $[0, 5]\times\left(R_\mathrm{planet}/R_\star\right)^2$ to prevent unphysical excursions at wavelengths where strong telluric absorption leaves no usable signal (see Equation~\ref{eq: clamps} and the discussion following it). The procedure then repeats until the mean absolute change of both the stellar flux and the telluric optical depth between consecutive iterations drops below a preset tolerance ($10^{-4}$ for the results presented in this paper), or until the maximum allowed number of iterations is reached.

\subsection{Degeneracies and uniqueness}\label{sec:forward_model:degeneracies}

There are some degeneracies between unknown vectors, even though they appear orthogonal in the data space (wavelength versus time or phase). The obvious one is between the scaling vector $S$ and the stellar flux $F_\lambda^\star$. This degeneracy is removed by fixing the value of $S$ at one particular phase (we keep $S_j$ fixed at its initial value of 1 at the phase with the lowest airmass). Anchoring $S$ also removes the ``continuum'' telluric opacity (a non-zero optical depth $\tau_\lambda$ present at all wavelengths), which can otherwise be imitated by an airmass-modulated variation of $S_j$.

A more subtle degeneracy affects the broadband level of the planet blocking area $P$. That level
enters the model only through the contrast between the in-transit and out-of-transit brightness, and the same
contrast can be produced by small compensating adjustments of the per-phase scaling $S_j$, of the level of the
stellar flux, or of per-night constant offsets of the telluric optical depth. The last channel is particularly
efficient when the airmass reaches its minimum near mid-transit, since a constant change of $\tau_\lambda$ then
produces a smooth brightness variation peaking at mid-transit that mimics a change of transit depth. Numerical
experiments confirm the analysis: solutions started from different mean levels of $P$ reach the same fit quality
and the same spectral shape, and the recovered level varies from one spectral segment to another by far more
than its statistical uncertainty (Section~\ref{sec:uncertainty}).

We do not impose the photometric transit depth on $P$. Fixing the mean level externally would prevent
the inversion from being tested against the photometry, and leaving it free costs nothing in fit quality. The
comparison is reported in Section~\ref{sec:uncertainty}: the median recovered level over the eighteen spectral
segments agrees with the photometric transit depth to three per cent, while the individual segments scatter by
far more than their statistical uncertainties. Photometry therefore remains the appropriate source of the
absolute level should one be required; what TSD measures is the spectral structure of the planetary
transmission.

A separate difficulty
arises in the cores of saturated telluric lines, which change little with airmass: there the only recourse is to
plan the observations so that the barycentric velocity ``scans'' the blocked regions, as the five additional
transits of Figure~\ref{fig:sim spectra for 7 transits} illustrate.

The uniqueness of the solution deserves a definition. In the strict mathematical sense a unique solution
would be a single global minimum of the discrepancy function, and we make no such claim: the observations carry
noise, so any reconstruction is a statistical estimate, and the separation of the components relies on a
diversity of Doppler shifts that is set by the observing time awarded rather than chosen. In this paper
uniqueness therefore has a practical meaning --- the reconstruction places the planetary features at the correct
wavelengths, does not, with few exceptions, produce features absent from the input, and recovers their relative
strengths approximately. The absolute level is the separate matter treated above. These are the properties
verified in Section~\ref{sec:simulator:performance}, both by direct comparison with the input transmission
spectrum and by cross-correlation.

TSD is working with the pixel sampling of observations but interpolations are unavoidable, which may lead to growing correlated noise.
To avoid that we use smoothing regularization for the stellar flux, the update of the
telluric absorption, and the planetary transmission. For each of the three spectral
unknowns the penalty added to the discrepancy function is a first-difference (gradient)
term,

\begin{equation}
\lambda\sum_i \left(x_i-x_{i-1}\right)^2 ,
\label{eq: regularization}
\end{equation}

the sum running over neighbouring pixels within a spectral segment. Its contribution to
the normal equations is $\lambda\,\mathsf{D}^{\mathsf{T}}\mathsf{D}$, with $\mathsf{D}$
the first-difference matrix; this is tridiagonal, adding $2\lambda$ to the diagonal at
interior pixels, $\lambda$ at the two end pixels, and $-\lambda$ to both off-diagonals.
There is no wrap-around and no reflection: the smaller weight at the ends is simply what
$\mathsf{D}^{\mathsf{T}}\mathsf{D}$ gives when no difference is defined across the segment
boundary. No penalty is applied in the phase direction and none couples different segments.
With the regularization terms included, Equations~\ref{eq: summary flux},
\ref{eq: summary tau} and~\ref{eq: summary P} become tridiagonal systems of linear
equations.

The three penalties are normalized differently, which is why the nominal parameters differ
by orders of magnitude. For the stellar flux the parameter is multiplied by the median of
the diagonal of the corresponding normal-equation matrix, so that the penalty is a fixed
fraction of the local data weight and is independent of the flux units. For the telluric
optical depth, which is dimensionless, the parameter is used directly. For the planetary
transmission it is referred to the photometric transit depth $(R_\mathrm{p}/R_\star)^2$
rather than to the current mean of $P$; the latter is altered by the bounds imposed below,
which can only raise values, so referring the penalty to it would let the objective drift
from one iteration to the next.

A quadratic penalty of this kind cannot enforce positivity. Positivity and the physical
bounds are imposed separately, after each update, by projection onto the feasible interval,

\begin{equation}
\tau \leftarrow \min\left(\max(\tau,0),\,\tau_\mathrm{max}\right), \quad
P \leftarrow \min\left(\max(P,0),\,P_\mathrm{max}\right), \quad
F \leftarrow \max(F,0) ,
\label{eq: clamps}
\end{equation}

with $\tau_\mathrm{max}=20$ and $P_\mathrm{max}=5\,(R_\mathrm{p}/R_\star)^2$. These
projections are not part of the minimized functional, and it is worth being explicit about
the consequence: a projection is not a descent step, so pixels that repeatedly reach a
bound bias the solution slightly at every iteration. A lower bound on $P$ placed inside the
noise distribution does exactly this, which is why the lower bound is zero rather than a
fraction of the transit depth. To verify how strongly the recovered transmission spectrum depends on the initial guess and regularization, we experimented with different initial guesses and regularization parameter values. The results are discussed at the end of Section~\ref{sec:simulator:performance}.

\subsection{Implementation}\label{sec:forward_model:implementation}

The algorithm described in the previous section was implemented using the IDL programming language, with a Python port being currently in development. The IDL code consists of the data input part, the initialization (initial guesses for all unknown functions) and the main loop that updates one unknown function at a time. The main loop calls sequentially four solvers (for stellar flux, telluric optical depth, scaling/offset, and planet blocking area) and it ends when the maximum number of iterations or convergence is achieved. The solver for each unknown function can be run with a diagnostic switch reporting the discrepancy function immediately before and after its update. Since each update is obtained by setting the corresponding partial derivative of $\Phi$ to zero, the discrepancy cannot increase across it, so this provides a direct check on both the derivation and its implementation; the analytic derivatives were in addition compared against numerical derivatives of $\Phi$ while the code was being written. The input data consists of observations grouped in individual transits, spectral intervals (e.g. from echelle orders or detectors) and orbital phases. As previously stated, we assume constant stellar flux and planet blocking area between transits but telluric absorption and observing conditions can change.

Observations are complemented by planetary and barycentric RVs used to compute the Doppler factors $\gamma$, $\beta$, and $\delta$ as well as the planet shadow location with respect to the stellar center (for computing blocked fraction of the stellar flux). Throughout the computation, the observations and other functions of wavelength need to be interpolated onto a different wavelength grid due to Doppler shifts. This is done using cubic splines and we avoid extrapolation. Only the parts of relevant vectors that can be evaluated without extrapolation are used by the solvers. As the result, the ends of spectral intervals are less constrained by the observations and rely more on regularization. The fact that we process one unknown function at a time slows somewhat the convergence: 300 iterations over all three detectors take 25 minutes on a modern desktop computer (Apple M2 Ultra CPU) using 114 phases and 36000 wavelength points. The number of iterations actually required is set by the tolerance adopted for the stopping criterion rather than by the difficulty of the minimisation.

The implementation of the TSD algorithm described above was tested using simulated data. To achieve this, we have developed a sophisticated simulation tool, capable of generating high-resolution ``observed'' spectra given the stellar flux spectrum, stellar specific intensities, the telluric absorption and the planetary transmission spectrum. The transit-specific parameters and instrument signatures can be replicated using real transits of real exoplanets and existing instruments. The output of the simulator has the same structure as the real data preparation tool: a single FITS file containing all the necessary information for the TSD code. The specifics of this simulator are described in the next section.

\section{TSD Performance on Simulated Data}\label{sec:simulator}

Alongside the development of the TSD, we developed software\footnote{Python code for interfacing with data reduction software, handing of template spectra, our exoplanet transmission spectroscopy simulator, and observational data cleaning and preparation can be found at \url{https://github.com/adrains/luciferase}.} to produce simulated exoplanet transmission spectroscopy observations. The combination of having a known ground ``truth'' for such simulations and the ability to generate simulations with several transits and arbitrary S/N values---something not shared by real data---was essential for testing and benchmarking TSD during its development. For comparison purposes, we first simulated two real transits observed by VLT/CRIRES+ (i.e. matching the exposure midpoints and durations, airmasses, and barycentric velocities of the real data) and an additional five real, but unobserved, transits. In order to run, the simulator requires template stellar, telluric, and exoplanet spectra; an adopted instrumental setup including spectral resolution and wavelength scale; data uncertainties, an exoplanet orbital solution and associated velocities. Subsection \ref{sec:simulator:system_info} describes our test case exoplanet system WASP-107 b, adopted system parameters, and model stellar and exoplanet spectra; Subsection \ref{sec:simulator:molecfit} describes our adopted model telluric spectra; Subsection \ref{sec:simulator:methodology} describes the simulation methodology itself; and finally, Subsection \ref{sec:simulator:performance} shows results and performance of the new algorithm when tested on simulated transmission spectroscopic data.

\subsection{The WASP-107 b System}\label{sec:simulator:system_info}
WASP-107 b is a warm super-Neptune orbiting a late K dwarf of spectral type K6, discovered by the WASP (Wide Angle Search for Planets, \citealt{2006PASP..118.1407P}) ground-based transit survey in 2017 \citep{anderson_discoveries_2017}. With a mass of only 0.12$\,M_{\rm Jup}$ but a radius nearly that of Jupiter at 0.94$\,R_{\rm Jup}$, its density is remarkably low and its transit depth is 2.17\%, which together with the short orbital period of 5.7$\,$days makes it highly amenable to atmospheric characterization at both low and high spectral resolution. It has an extended helium-rich atmosphere and a comet-like helium tail \citep{spake_wasp107_2018, allart_high-resolution_2019, kirk_confirmation_2020, spake_posttransit_2021, guilluy_gaps_2024}, and previous studies have detected H$_2$O, CH$_4$, CO, CO$_2$, SO$_2$ and NH$_3$ \citep{kreidberg_wasp107_2018, dyrek_wasp107_2024, welbanks_wasp107_2024}.

The planet follows a misaligned, nearly polar orbit; \citet{rubenzahl_tess-keck_2021} demonstrated its retrograde motion from the Rossiter-McLaughlin \citep[RM,][]{1924ApJ....60...15R,1924ApJ....60...22M} effect measured with Keck HIRES. The RM effect also measures the stellar $v\sin i$ directly rather than bounding it, and they obtained 0.45\,\kms, compatible with the limb-darkening approximation introduced in Section~\ref{sec:forward_model:specific intensity}. Both the low $v\sin i$ and the high obliquity are advantageous for high-resolution transmission spectroscopy, each reducing the significance of the RM signature in our observations and simulations.

Its puffy atmosphere, favourable orbital geometry, confirmed atmospheric detections and slowly rotating host make it an excellent test system for demonstrating TSD. Characterization of this planet at high spectral resolution in the infrared is nevertheless a relatively unexplored regime. \citet{Linn_WASP107_2025} analysed the very same VLT/CRIRES+ observations used here (Section \ref{sec:spectra}) with SYSREM and cross-correlation, and found the target challenging in the $K$ band from the ground, largely because of tellurics and a high-altitude cloud deck that quenches the spectral features; they argue that the same difficulties should affect other cool, cloudy planets rather than hot Jupiters. That conclusion is among the motivations for the present work.

Sections \ref{sec:simulator:system_info:star} and \ref{sec:simulator:system_info:planet} describe synthetic stellar and planet transmission spectra used in our simulations. The adopted parameters are listed in Table \ref{tab:syst_info}; we confirmed the stellar parameters with SME \citep{SME2} using ESO HARPS observations of WASP-107 from 2018 (programme 0100.C-0750(D)).

\begin{table}
\centering
\caption{WASP-107 system stellar, planet and orbital parameters. 
}
\label{tab:syst_info}
\begin{tabular}{cccc}
\hline
Parameter & Unit & Value & Reference \\
\hline
\multicolumn{4}{c}{Star} \\
\hline
Gaia DR3 ID & - & 3578638842054261248 & \citet{vallenari_gaia_2023} \\
2MASS ID & - & 2MASS J12333284-1008461 & \citet{skrutskie_two_2006} \\
RA & hh:mm:ss.ss & 12 33 32.74 & \citet{vallenari_gaia_2023} \\
Dec & dd:mm:ss.ss & -11 51 13.62 & \citet{vallenari_gaia_2023} \\
Parallax & mas & $15.5277\pm0.0260$ & \citet{vallenari_gaia_2023} \\
Distance & pc & $64.40\pm0.03$ & \citet{vallenari_gaia_2023} \\
RV & km\,s$^{-1}$ & $14.06\pm0.20$ & \citet{vallenari_gaia_2023} \\
$K_S$ mag & - & $8.637\pm0.023$ & \citet{skrutskie_two_2006} \\
$M_\star$ & $M_\odot$ & $0.683\pm0.017$ & \citet{piaulet_wasp-107bs_2021} \\
$R_\star$ & $R_\odot$ & $0.67\pm0.02$ & \citet{piaulet_wasp-107bs_2021} \\
$T_{\rm eff}$ & K & $4425\pm70$ & \citet{piaulet_wasp-107bs_2021} \\
$v \sin i$ & km\,s$^{-1}$ & $2.5\pm0.8$ & \citet{anderson_discoveries_2017} \\
\hline
\multicolumn{4}{c}{Planet} \\
\hline
$M_P$ & $M_\oplus$ & $30.5\pm1.7$ & \citet{piaulet_wasp-107bs_2021} \\
$R_P$ & $R_\oplus$ & $10.04\pm0.2$ & \citet{mocnik_starspots_2017} \\
$a$ & AU & $0.0553\pm0.0013$ & \citet{mocnik_starspots_2017} \\
$e$ & - & $0.06\pm0.04$ & \citet{piaulet_wasp-107bs_2021} \\
$i$ & deg & $89.56\pm0.078$ & \citet{mocnik_starspots_2017} \\
$K$ & m\,s$^{-1}$ & $14.1\pm0.8$ & \citet{skrutskie_two_2006} \\
Transit Duration & hr & $2.73864\pm0.00194$ & \citet{mocnik_starspots_2017} \\
JD (mid) & day & $2457584.329746\pm0.000011$ & \citet{mocnik_starspots_2017} \\
$P$ & day & $5.7214742$ & \citet{piaulet_wasp-107bs_2021} \\
$T_\mathrm{eqil}$ & K & $672.5\pm147.5$ & \citet{2025ApJS..278...52H} \\
\hline
\end{tabular}
\end{table}\label{System Info}

\subsubsection{WASP-107 Stellar Spectrum}\label{sec:simulator:system_info:star}

We computed the stellar spectrum using a 1D LTE\footnote{Local Thermodynamic Equilibrium} spherically symmetric MARCS model atmosphere \citep{gustafsson_grid_2008} and the associated \texttt{BSYN} radiative transfer software at R${\sim}200\,000$ for $T_{\rm eff}=4\,420\,$K, $\log g=4.61$, ${\rm[M/H]}=0.05$. This version of MARCS adopts solar abundances by \citet{grevesse_solar_2007}, [$\alpha$/Fe] scaled by the metallicity [Fe/H], line data per the VALD-2 database \citep{stempels_vald_2012}, and opacity tables per references in \citet{gustafsson_grid_2008}. The use of the spherically symmetric model atmosphere enables more accurate computation of specific intensity spectra at the starting and ending phases of the transit. For this work, we sample 49 different $\mu$ angles across the disk of the star with density increasing toward the limb, and with uniform sampling in wavelength. This is sufficient to perform interpolation of specific intensities to the position of the planet throughout the transit, \textit{without} involving any approximation for limb darkening.

\subsubsection{WASP-107 b Transmission Spectrum}\label{sec:simulator:system_info:planet}
We generated the template exoplanet spectra using the Python radiative transfer package \texttt{petitRADTRANS} version 2.7.6\footnote{\url{https://petitradtrans.readthedocs.io/}} \citep{molliere_petitradtrans_2019}. We included absorption from H$_2$O \citep{Exomol_Polyansky_2018}, CO \citep{hitemp_rothman_2010}, NH$_3$ \citep{exomol_yurchenko_2011} based on the analysis of WASP-107 b in \cite{Linn_WASP107_2025}. The templates were initially generated at a spectral resolution of $R=10^6$ over the 1.6–3.0 $\mu$m wavelength range, and later convolved and clipped to match the nominal resolving power ($R$ = 100,000) and wavelength range (2.0–2.5 $\mu$m) of the real CRIRES+ $K$ band observations. The assumed parameters for the planetary atmosphere came primarily from the recent JWST studies of WASP-107 b such as  \citet{dyrek_wasp107_2024}, \citet{welbanks_wasp107_2024}, and \citet{Sing_2024_wasp} including $T_{\mathrm{int}}$ = 460 K, $T_{\mathrm{eq}}$ = 738 K, and $R_{\mathrm{p}} = 0.94 \, R_{\mathrm{J}}$. A complete list of input parameters in simulating the exoplanet transmission spectrum can be found in Table 3 of \cite{Linn_WASP107_2025}.

\subsubsection{\texttt{Molecfit} Telluric Model}\label{sec:simulator:molecfit}
To generate a template telluric spectrum for our simulations, we make use of the \texttt{Molecfit} software package \citep{smette_molecfit_2015, kausch_molecfit_2015}, which was developed as a model-based alternative to the traditional---but observationally expensive---approach of observing telluric reference stars interleaved with science targets over a night. \texttt{Molecfit} is designed to remove telluric features from science observations, and it does this by computing an initial guess taking into account the real atmospheric profile at the telescope location at the time of the science observations and morphs it to find the best match to predicted telluric features by adjusting abundances, broadening, and additional parameters. The synthetic telluric spectrum is computed by solving radiative transfer line-by-line using a large collection of data for all molecular absorbers present in the Earth atmosphere (including H$_2$O, CO, CO$_2$, CH$_4$---the main species affecting the $K$ band).

We require one unique telluric transmission spectrum per simulated night. To that end, we use \texttt{Molecfit} to fit the exposures of seven real VLT/CRIRES+ transits observed in the \texttt{K2148} setting: two actually of WASP-107 b, and in lieu of additional WASP-107 b observations, another five transits of GJ 1214 b, all observed as part of Guaranteed Time Observations (GTO) for CRIRES+. To improve the fit, these observations were pre-processed as described in Section \ref{sec:spectra:reductions} in order to reduce the reliance on \texttt{Molecfit}'s own continuum correction capability. Starting with blaze-correction of CRIRES+ spectra, continuum normalization was done separately for each spectral order and each of the three detectors using a linear polynomial fit. The `continuum' points were selected by eye and strong stellar lines identified by comparison against the synthetic stellar spectrum were masked out as explained in Section \ref{sec:simulator:system_info:star}. \texttt{Molecfit} was used in a two-step iterative process to improve the final telluric fit. After the first \texttt{Molecfit} iteration, we tune the continuum polynomial with respect to the CRIRES+ observations, synthetic stellar spectrum, and the initial telluric model. Then we run \texttt{Molecfit} a second time to produce the final adopted telluric model for the night in question. \texttt{Molecfit} is run on every exposure, and returns a transmission spectrum at the airmass of that exposure, whereas the simulator needs the zenith optical depth of Equation~\ref{eq: forward model}. Each fit is therefore converted to the optical depth by removing the airmass factor: $\tau_\lambda = -\ln T_\lambda / z_j$, and the median over the exposures for each transit is adopted as the optical-depth vector of that transit. We use median rather than the mean because \texttt{Molecfit} occasionally fails on an individual exposure. For our 7-transit simulations the $\tau$-vectors were obtained from the two WASP-107 b transits (first two transits) and the five GJ 1214 b transits nights. Thus every simulated transit carries the telluric conditions of a different real night at Paranal.

\begin{figure*}[hbt!]
    \centering
    \includegraphics[width=0.8\textwidth, trim=2.5cm 8.3cm 2.5cm 10.cm]{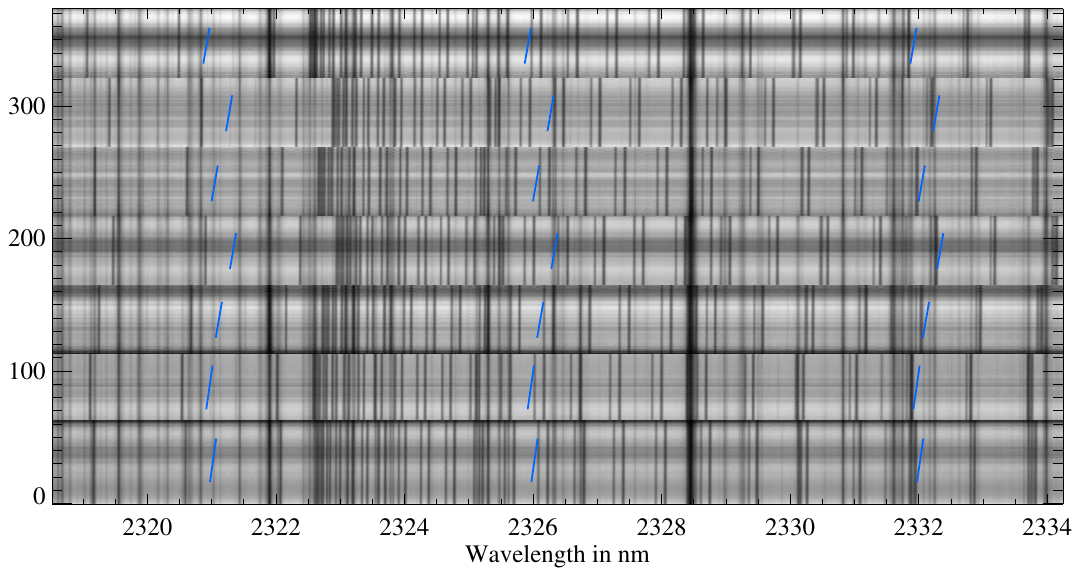}
    \caption{A wavelength fragment of the simulated seven-transit observations of WASP-107 b with VLT/CRIRES+. The wavelength range (horizontal axis, nm) corresponds to the part of spectral order 24 (second order from the bottom) that falls onto the middle detector, and the vertical axis is the time dimension starting with exposure 0 of the first observation at the bottom. Horizontal bands correspond to individual transits with 50-64 spectra per transit, with the influence of airmass and variable slit losses affecting the flux level between adjacent exposures. Blue lines indicate how selected wavelengths, fixed in planetary restframe will shift in observations. One can clearly distinguish telluric (constant in wavelength), stellar (shifting with barycentric velocity) and planetary (shifting and tilted) lines. Notice the variation of telluric absorption strength between transits.}
    \label{fig:sim spectra for 7 transits}
\end{figure*}

\subsection{Simulation Methodology}\label{sec:simulator:methodology}
Our exoplanet transmission spectroscopy simulations consist of two main steps. The first involves the compilation of the so-called `time-step info' for each exoplanet phase (i.e. observed spectrum) for each night to be modeled. This includes compiling self-consistent timing, positional, and velocity information from the real observations (including number of exposures, exposure times, airmasses, barycentric velocities), as well as computing orbital geometry of the exoplanet itself (e.g. projected planet positions and velocities, orbital phase). The second step involves the simulator proceeding nightly to model each phase via Equation \ref{eq: forward model} using our template stellar, telluric, and exoplanet spectra, plus the per-phase time-step information. Critically, we model the \textit{extracted} 1D spectra for each phase, and do not simulate individual optical elements, detectors, telescope nodding, or data-reduction---instead approximating these as a series of transfer functions. We compute and interpolate each component of the equation separately---disk integrated flux (RV frame $\gamma$), \change{occulted stellar specific intensity (RV frame $\beta$), exoplanet effective blocking area (RV frame $\delta$)}, airmass-scaled telluric transmission, and scaling value representing the slit losses.\footnote{Note that we do not simulate the offset term $D_j$ from Equation \ref{eq: forward model}. The extra background offset was a provision to account for imperfect nodding pair subtraction. When applied to the real data, it turned out to be insignificant (less than 5 photoelectrons).}
Finally, the spectra are convolved with the instrumental transfer function, and a realistic noise pattern is added. The noise is drawn per pixel from the uncertainties reported by the CRIRES+ data reduction system for the corresponding real exposure, so the signal-to-noise ratio varies with wavelength and with phase rather than taking a single value; the figure of 130 quoted here and below is a representative continuum value for one spectrum and not the noise model itself. These uncertainties are reliable except in the cores of the strongest telluric features, where almost no flux is transmitted, but those pixels do not determine the solution: the attenuation that makes their uncertainties unreliable also makes them less important, as noted in Section~\ref{sec:forward_model:minimisation}. The S/N of 130 and slit losses were estimated using the real observations of WASP-107 b with CRIRES+ analyzed in \citet{Linn_WASP107_2025} and in Section \ref{sec:spectra}. The interpolation to different restframes is performed with cubic spline using the \texttt{interp1d} function of Python's \texttt{scipy}.

It is worth being explicit about the extent to which these simulations constitute an independent test of the inverse method. The simulator and the inversion share the structure of Equation~\ref{eq: forward model}: the decomposition into stellar, planetary, and telluric components, the assumption that the stellar and planetary spectra are constant in time, and the representation of telluric absorption by a single optical-depth spectrum per night scaled by the airmass. Several other elements are, however, not shared between the two. The specific intensities are obtained by spectral synthesis at 49 limb angles (Section~\ref{sec:simulator:system_info:star}), whereas the inversion uses the analytical limb-darkening relation derived in Section~\ref{sec:forward_model:specific intensity}. The planetary transmission spectrum comes from a radiative-transfer calculation with \texttt{petitRADTRANS}, and the telluric spectra are \texttt{Molecfit} fits to real CRIRES+ observations, so both carry structure that the forward model of the inversion cannot generate. All components are computed over a wider wavelength range and at a resolution of $R\gtrsim200\,000$, the instrumental profile is applied to their \textit{product} of Doppler-shifted components rather than to each component separately, and the result is then resampled onto the wavelength grid of the pipeline. The tests presented below therefore do probe the accuracy of the limb-darkening approximation, the commutation of the instrumental convolution with the multiplicative combination of the components, and the effect of the spline resampling---none of which is exact in the inverse model---while leaving the assumptions listed above to be examined with real data.

The result is one datacube of fluxes and uncertainties per transit of shape $[N_{\rm phase}, N_{\rm spec}, N_{\rm px}]$, alongside a table of time-step information of length $N_{\rm phase}$, where $N_{\rm phase}$ is the number of phases, modelled for this night; $N_{\rm spec}$ is the number of spectral segments (three detectors per spectral order for CRIRES+); and $N_{\rm px}$ is the number of spectral pixels per spectral segment. The spectra and time-step information for all nights simulated are then packaged in the same custom FITS file format as the real data, allowing TSD to be agnostic as to whether it is running on real or simulated data.

\begin{figure*}[htb!]
    \centering
    \includegraphics[width=0.93\textwidth, trim=2.5cm 4cm 1.5cm 3.7cm]{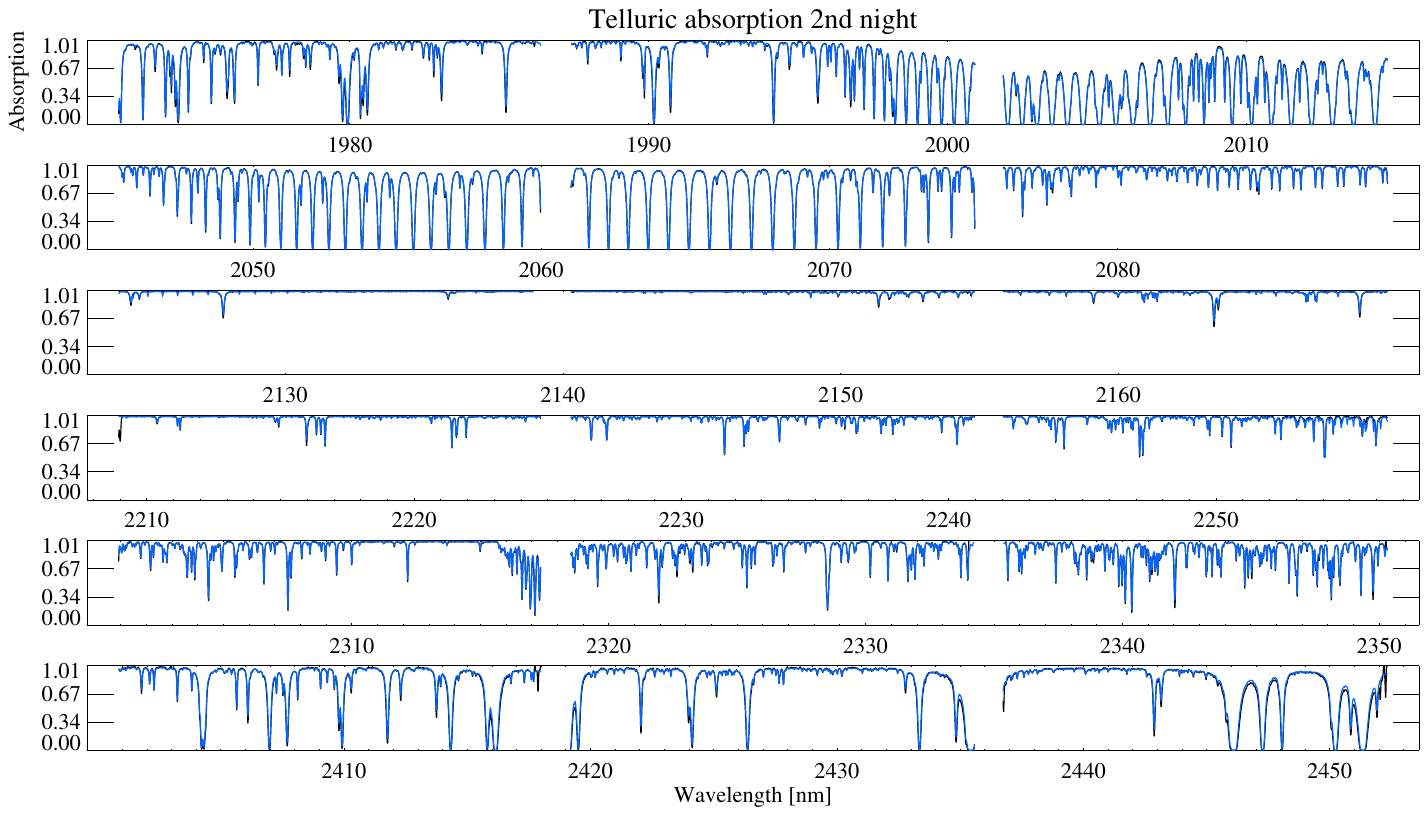}
    \includegraphics[width=0.93\textwidth, trim=2.5cm 3.5cm 1.5cm 3.7cm]{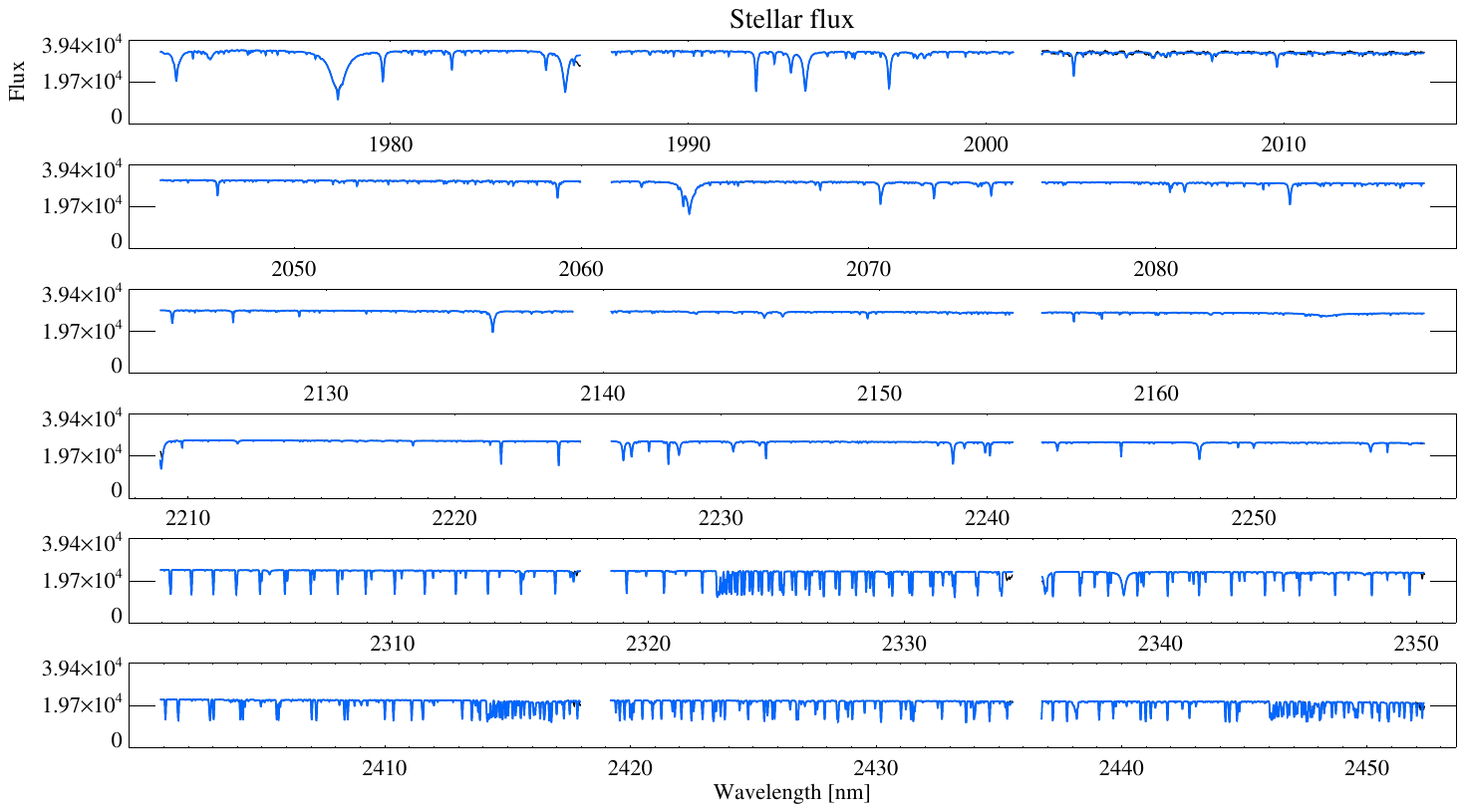}
    \caption{\textbf{Top}: Reconstruction of the telluric absorption from the simulated seven transit observations of WASP-107 b (in black) for the second transit. 6 spectral orders registered by CRIRES+ in K band are shown vertically with 3 horizontal segments corresponding to CRIRES+ detectors. The true solution used in the simulations (without noise) is shown in blue. Note that TSD solves for optical thickness $\tau$ and what is plotted is the telluric absorption in zenith $e^{-\tau}$. \textbf{Bottom}: As above, but for the reconstruction of the stellar flux.}
    \label{fig:telluric and flux reconstruction from 7n sim}
\end{figure*}
\begin{figure*}[htb!]
    \centering
    \includegraphics[width=\textwidth, trim=1.5cm 1.7cm 1.5cm 1.5cm]{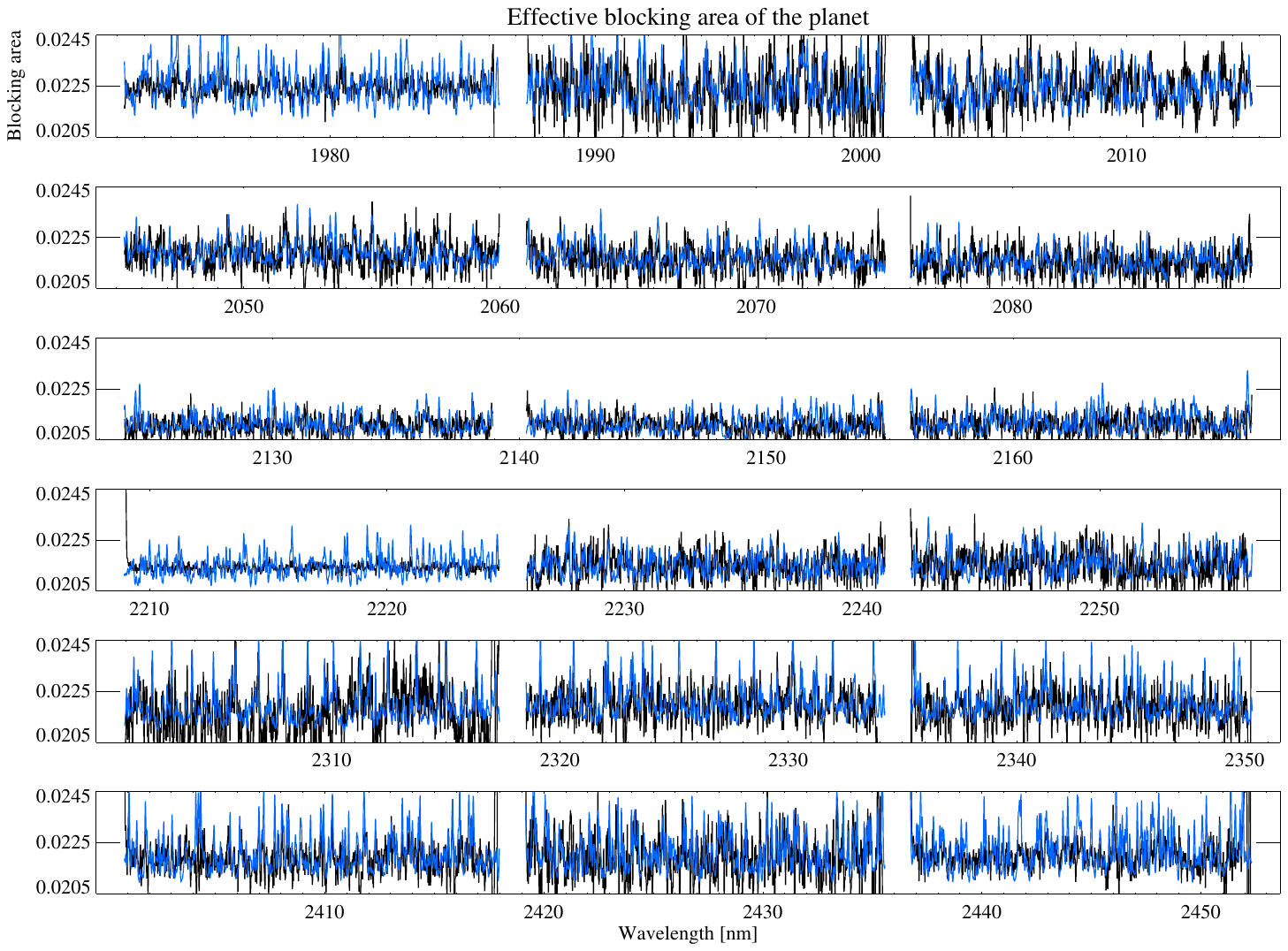}
    \caption{As Figure \ref{fig:telluric and flux reconstruction from 7n sim}, but for the planetary component of WASP-107 b. The black line is the reconstruction of the planet effective blocking area based on seven simulated transits of WASP-107 b, and the true---and zero noise---solution that went into simulations is shown in blue. A zoom in view of the recovery for detector 2 is shown in Figure \ref {fig:trans reconstruction det2 from 7n sim}.}
    \label{fig:trans reconstruction from 7n sim}
\end{figure*}
\begin{figure*}[htb!]
    \centering
    \includegraphics[width=\textwidth, trim=1cm 1.7cm 1.5cm 1.5cm]{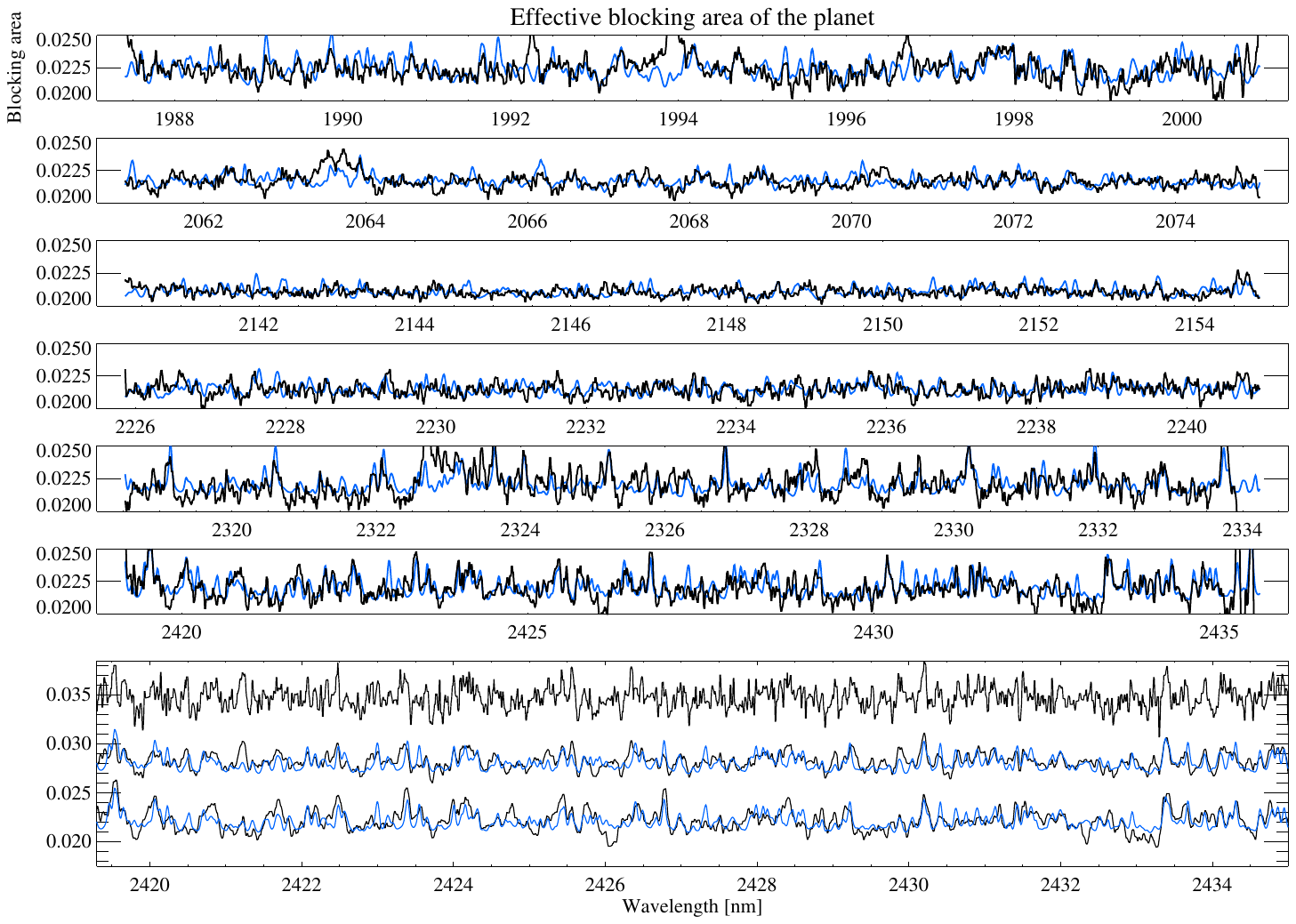}
    \caption{Zoom-in on the central detector (detector 2) of Figure~\ref{fig:trans reconstruction from 7n sim}. The top six panels show
    the TSD reconstruction in black and the true solution in blue. The bottom panel uses the reddest spectral interval to illustrate the
    sensitivity of the solution to the initial guess and the effect of regularization. We ran TSD twice. In the ``all tellurics'' case (bottom curve)
    the telluric absorption was set to
    the observation taken at the minimum airmass of each transit, while the stellar flux was given no spectral features. In the
    ``all stellar'' case (middle curve) the stellar flux was set to the observations at a randomly
    selected phase just outside the transit, while the telluric absorption was set to zero. The two reconstructions agree rather
    well. The top curve uses the same initial guess as the middle one but with the regularization parameter for the transmission spectrum
    reduced by a factor of ten relative to the default value. \change{The three curves are offset vertically for clarity.} Note that the wavelength range of the bottom panel is slightly reduced
    for better visibility.}
    \label{fig:trans reconstruction det2 from 7n sim}
\end{figure*}

\begin{figure*}[htb!]
    \centering
    \includegraphics[width=\textwidth, trim=1cm 1.7cm 1.5cm 1.5cm]{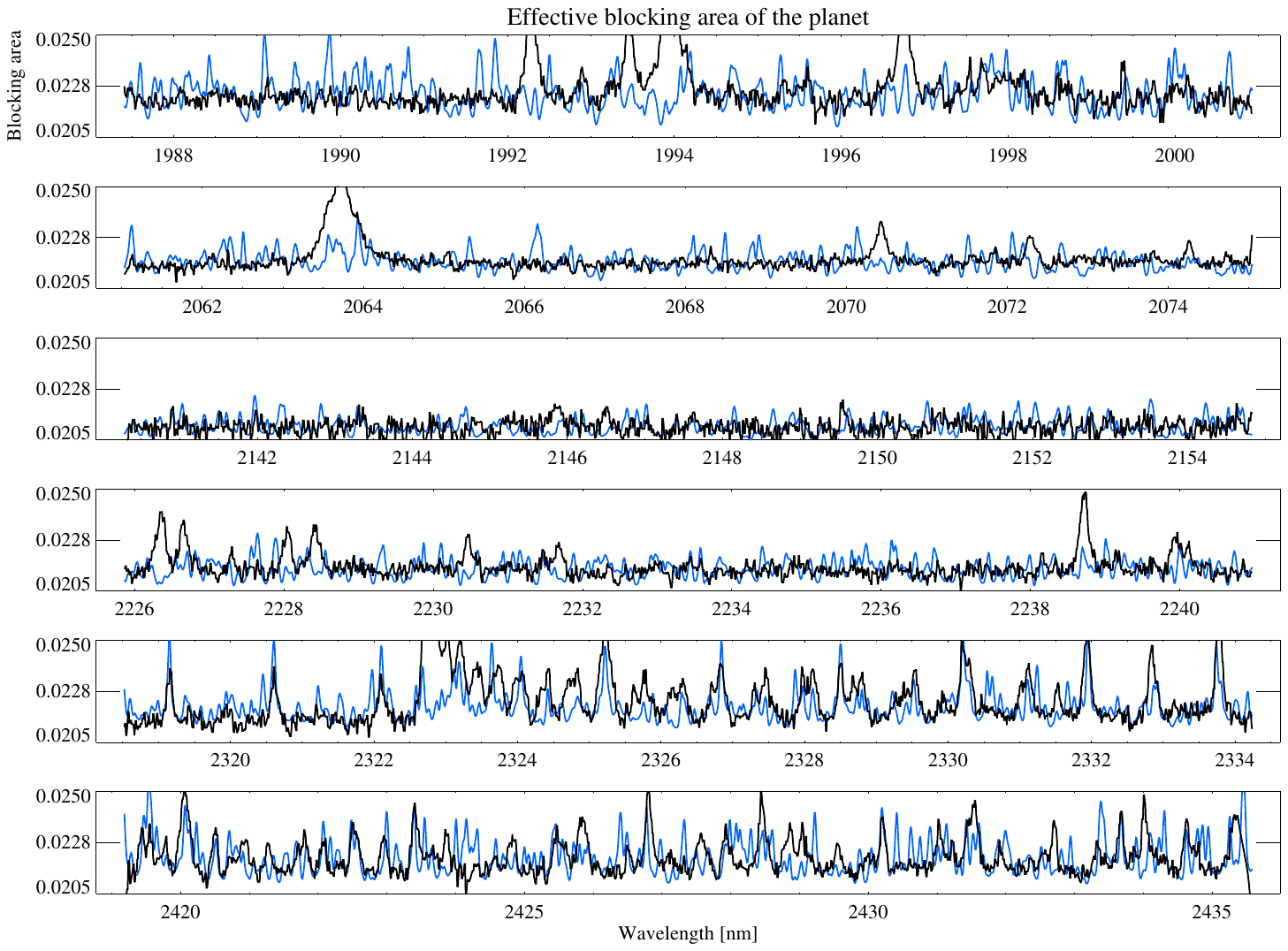}
    \caption{The same as in Figure~\ref{fig:trans reconstruction det2 from 7n sim} but based on the two first transits only. Note that regularization damps the small-scale features seen in the seven-night reconstruction, but it cannot remove large artifacts appearing due to the difficulty of disentangling spectral components given only two nights of data.}
    \label{fig:trans reconstruction det2 from 2n sim}
\end{figure*}

\begin{figure*}[htb!]
    \centering
    \includegraphics[width=0.4\textwidth, trim=3.4cm 2cm 2cm 1cm]{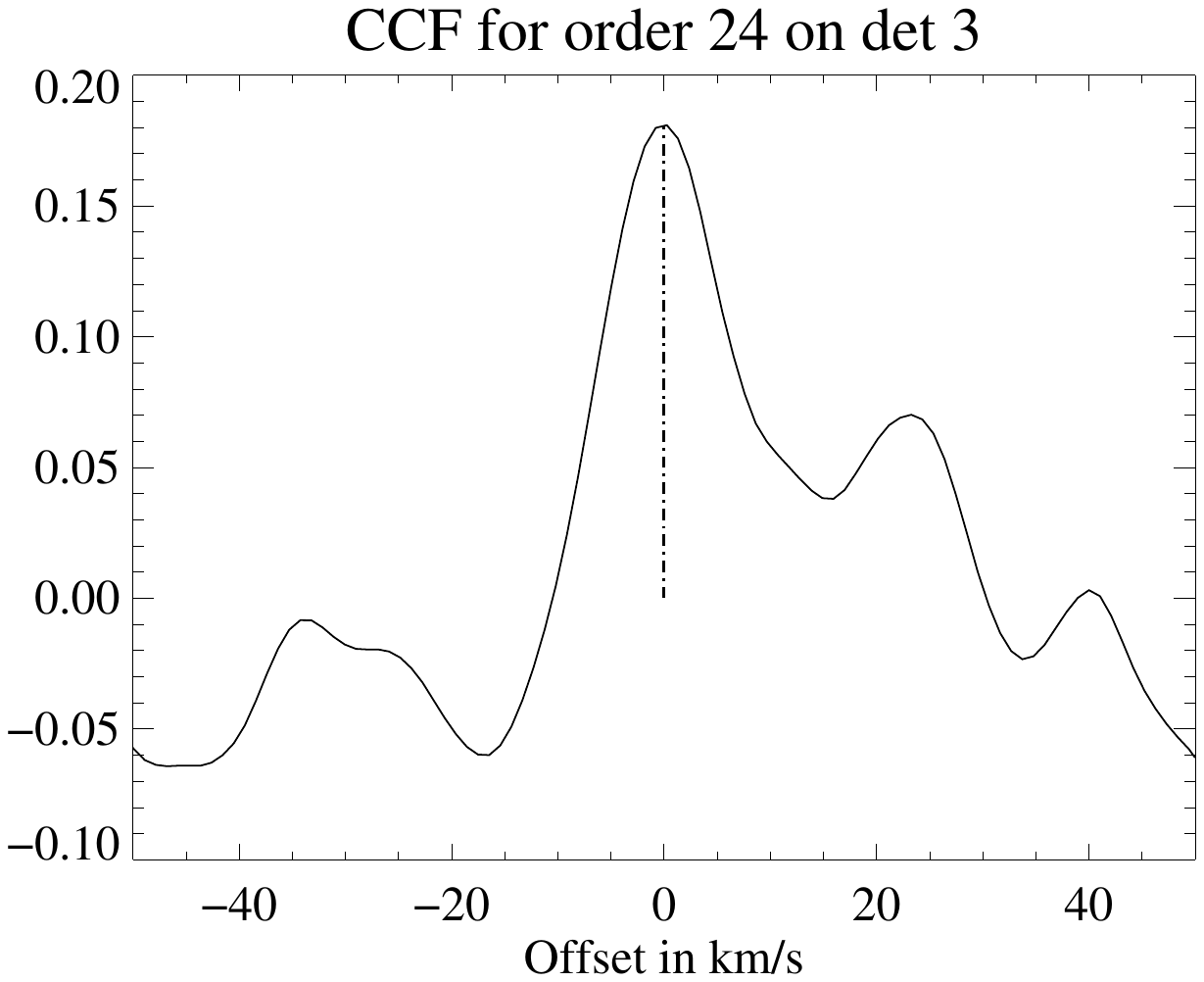}
    \includegraphics[width=0.58\textwidth, trim=0cm -0.5cm 0.2cm 0cm]{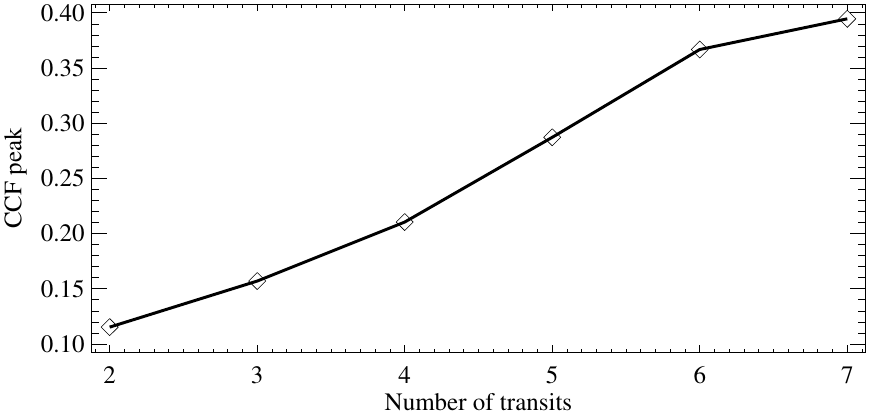}
    \caption{\textbf{Left}: an example of a CCF between the simulated and recovered transmission for a 7 transit simulation. The peak of the CCF (given in Table~\ref{tab:sim results}) is measured from 0 at zero velocity (shown by dash-dotted line). \textbf{Right}: CCFs peak values as a function of the number of transits given to the TSD (averaged for all spectral intervals). The value of the CCF peak increases with the number of transits, indicating how additional transits with different or `shuffled' barycentric velocities help to improve S/N and reduce degeneracies during the inversion procedure.}
    \label{fig:ccf versus number of transits}
\end{figure*}

\subsection{Model Performance}\label{sec:simulator:performance}
For testing the performance of the TSD we have produced simulated data for seven transits of WASP-107 b as described in Section~\ref{sec:simulator:system_info}, with a representative spectral order  shown in Figure~\ref{fig:sim spectra for 7 transits}. This illustrates the main reason TSD works: different transits separate or `shuffle' the stellar and telluric features because of the change of the barycentric velocity from night-to-night, while the orbital motion of the planet allows one to distinguish the planet transmission from the other two components. The latter, of course, requires high spectral resolution only possible from the ground.

The TSD inversion procedure was run on the first two, three, four, five, six, and then all seven simulated transits to test how the number of transits influences the quality of the recovery. Convergence was judged by the maximum change in the effective blocking area of the planet $P$ between successive iterations. Figure \ref{fig:telluric and flux reconstruction from 7n sim} shows the reconstruction when running on all seven simulated nights of (i) the telluric spectrum (night \#2 of 7) on the top, and (ii) the stellar flux on the bottom. Note that we choose to show telluric \textit{absorption} even though TSD works with the optical thickness because it is easier to see the interdependence between tellurics and other unknown functions in this way. The reconstruction (in black) is barely visible behind the input telluric absorption (in blue) and while occasional discrepancies are found at the ends of spectral segments, this is to be expected when avoiding extrapolation as we do. Generally, the reconstruction of tellurics and stellar flux is close to perfect considering that the S/N of an individual spectrum (i.e. a single phase) is around 130.

Finally, the transmission spectrum of the planet (the effective blocking area of the planet $P$ to be precise) based on seven transits is shown in Figure~\ref{fig:trans reconstruction from 7n sim}. The quality is obviously inferior to other unknown functions, but we have to remember that here the signal is at the level of 1000--5000 ppm of the stellar flux. This can be compared with the pressure scale height H in the atmosphere of WASP-107 b. We estimate H using the planetary parameters given in Table~\ref{System Info} and get a value of $630\pm185$~km. A change of the blocking radius of the planet by one scale height corresponds to the flux change of 380 ppm and so the spectral features we see in the transmission spectrum correspond to 5 and more  scale heights.

Figure~\ref{fig:trans reconstruction from 7n sim} allows us to draw the main conclusion that most of the transmission variations are well reproduced. Some discrepancies, such as at 1992 and 1994~nm, are associated with strong overlapping stellar and telluric features. The RV excursion of the transmission spectrum during transits does not exceed the width of these features (a few tens of \kms) and so the reconstruction fails. This is clearly a limitation of our method. One can also notice the ``edge effects" in several wavelength segments (e.g. on the right side of each segment in the two bottom orders). This is again related to the algorithm avoiding extrapolation of unknown vectors, and so not all phases contribute to these wavelengths when linearization matrices are computed.

The quality of the fit is measured by the reduced $\chi^2$,
\begin{equation}
\chi^2_\nu = \frac{1}{N_\mathrm{data}-N_\mathrm{param}}
             \sum_{\lambda,j}\left(\frac{M_\lambda^j-O_\lambda^j}{\sigma_\lambda^j}\right)^2 ,
\label{eq: reduced chi2}
\end{equation}
where $\sigma_\lambda^j$ are the per-pixel uncertainties delivered by the reduction pipeline (for the simulations, the uncertainties used to generate the noise). We note that these uncertainties are used only to evaluate the fit: the solution itself is obtained with uniform weights, as explained in Section~\ref{sec:forward_model:minimisation}. For the two transits of real data the observations comprise 4114260 points and
3965796 degrees of
freedom\footnote{114 phases $\times$ 18 spectral segments $\times$
2005 wavelength points = 4114260 data points. The unknowns are one stellar
flux and one transmission spectrum per spectral segment ($2\times18\times2005
= 72180$), one telluric spectrum per segment per transit
($2\times18\times2005 = 72180$), and a scaling factor and an additive offset
for every segment at every phase ($114\times18\times2 = 4104$), giving 148464
free parameters in total. The simulated segments are not trimmed and retain all
2048 pixels, so the corresponding two-transit simulation has 4202496 data points,
151560 unknowns and 4050936 degrees of freedom.}. For the simulated data TSD converges to $\chi^2_\nu$ between 0.6 and 1.2, depending on the number of transits, i.e. the reconstruction reproduces the simulated observations at the level of the injected noise, as it should. We stress that this is not a self-fulfilling test: as
described in Section~\ref{sec:simulator}, the simulated spectra were built
from specific intensities obtained by spectral synthesis, a
petitRADTRANS transmission spectrum and Molecfit telluric spectra, all at
higher resolution than the data and convolved with the instrumental profile
delivered by the pipeline. The simulator therefore does not share the forward
model of the solver, and agreement at the noise level is a statement about the
recovery rather than about a shared parametrisation.

For the real data (see Section~\ref{sec:spectra}) the pipeline uncertainties require care. At good signal levels they agree with the expected detector noise: with the gain of 2.15\,e$^-$/ADU of the instrument, the model $\sigma^2 = S/g + \mathrm{RN}^2$, where $S$ is the signal in ADU and $\mathrm{RN}\approx20$\,ADU is the effective read noise, reproduces the delivered uncertainties to better than 2\,\% at continuum level. Below 1000\,ADU, however, the pipeline estimates become less reliable and a small number of pixels are assigned uncertainties orders of magnitude smaller than the detector noise; these few pixels would dominate the sum in Equation~\ref{eq: reduced chi2} entirely. We therefore bound $\sigma_\lambda^j$ from below by the detector noise model quoted above, which leaves the well-measured pixels unchanged. The final solution then gives $\chi^2_\nu\approx1.21$ averaged over the three
detectors: the residuals exceed the photon and read noise by about 10\,\% in
amplitude per pixel. The value varies between detectors---1.41, 1.09
and 1.13 for the first, second and third respectively---and since the three
detectors are solved independently, this spread reflects differences in the
data rather than in the method. The remaining excess is not white noise but is concentrated in a small fraction of the pixels---0.4\,\% of them contribute 24\,\% of $\chi^2$, and 1\,\% of them 31\,\%---and most probably is associated with detector defects that survive the nodding-pair subtraction. Such effects are absent from the simulations by construction, and they, rather than the photon noise, set the current limit of the fit quality.

To examine recovery at the limit of our spectral resolution (3~\kms), we zoom in on the central detector in Figure~\ref{fig:trans reconstruction det2 from 7n sim}. The reconstruction looks good, but how good is it actually? Table~\ref{tab:sim results} presents the Root Mean Square (RMS) for each spectral interval, as well as the values of the peaks of cross-correlation between the input transmission that went into the simulations (planet blocking area) and its reconstruction (shown as blue and black lines in Figure~\ref{fig:trans reconstruction from 7n sim} for the seven-transitions simulation). Note that only the top two numbers in each cell of this table refer to the simulations, as the third number refers to the real data analyzed in Section \ref{sec:spectra}. The values in the table change significantly between spectral segments affected by the strength and number of spectral features present in each segment, but they give a good sense of improvement when increasing the number of transits.

The RMS and the CCF peaks were computed for each spectral segment separately. We restricted calculations to the central 2000 pixels in each segment because pixels in the ends of a segment are less constrained by the observations as they are Doppler-shifted in and out of the observed interval. The RMS and the CCF were computed as:
\begin{eqnarray}
    \mathrm{RMS}&=&\sqrt{\frac{1}{N_x}\sum_x \left(P_x^\mathrm{inp}-P_x^\mathrm{TSD}\right)^2} \\
    \mathrm{CCF(y)}&=&\frac{\sum_x \left(P_{x+y}^\mathrm{inp}-\bar{P}_{x+y}^\mathrm{inp}\right)
                             \cdot \left(P_x^\mathrm{TSD}-\bar{P}_x^\mathrm{TSD}\right)}
                     {\sqrt{\sum_x \left(P_x^\mathrm{inp}-\bar{P}_{x}^\mathrm{inp}\right)^2}
                \cdot \sqrt{\sum_x \left(P_x^\mathrm{TSD}-\bar{P}_x^\mathrm{TSD}\right)^2}},
\end{eqnarray}
where $x$ is the element number of the transmission vector (wavelength) and $\bar{P}$ is the mean value of the corresponding vector. For the CCF, the range of offsets given by $y$ was selected to be $\pm 50$ \kms. The peak value of the CCF was measured from zero at the zero offset as illustrated in the left panel of Figure~\ref{fig:ccf versus number of transits}.

\begin{table}
\centering
\caption{Numerical assessment of the reconstruction results for simulations with two and seven transits as well as for the two-transit real data of WASP-107 b. The format of the table is similar to the layout of Figures~\ref{fig:telluric and flux reconstruction from 7n sim} and \ref{fig:trans reconstruction from 7n sim}. The rows corresponds to spectral order numbers, and the columns to detectors. Each cell contains a stack of three numbers: \change{RMS or cross-correlation peak of the simulated versus recovered transmission spectrum for two (top), seven (in bold, in the middle) and for the real data (italic, bottom). Note that a direct comparison between recovered and real transmission is not possible and so we use petitRADTRANS-simulated proxy to derive the bottom numbers.}}
\label{tab:sim results}
\begin{tabular}{lllllll}
\hline
Order &\multicolumn{2}{c}{Det 1} & \multicolumn{2}{c}{Det 2} & \multicolumn{2}{c}{Det 3} \\
   &   RMS   &  CC    &   RMS   &   CC   &   RMS   &   CC   \\    
\hline
28 & 0.00102 & 0.0498 & 0.00102 & 0.0178 & 0.00072 & 0.1775 \\
   & \textbf{0.00096} & \textbf{0.1551} & \textbf{0.00085} & \textbf{0.2985} & \textbf{0.00073} & \textbf{0.1935} \\
   & \textit{0.00136} & \textit{0.0841} & \textit{0.00110} & \textit{0.0190} & \textit{0.00112} & \textit{0.0701} \\   
\hline
27 & 0.00061 & 0.0522 & 0.00054 & 0.1858 & 0.00051 & 0.1157 \\
   & \textbf{0.00051} & \textbf{0.3264} & \textbf{0.00049} & \textbf{0.3043} & \textbf{0.00046} & \textbf{0.2871} \\
   & \textit{0.00074} & \textit{0.0414} & \textit{0.00128} & \textit{0.0860} & \textit{0.00069} & \textit{0.0782} \\
\hline
26 & 0.00047 & 0.0941 & 0.00046 & 0.1203 & 0.00050 & 0.1957 \\
   & \textbf{0.00043} & \textbf{0.2496} & \textbf{0.00044} & \textbf{0.2106} & \textbf{0.00048} & \textbf{0.2519} \\
   & \textit{0.00050} & \textit{0.0205} & \textit{0.00067} & \textit{0.1075} & \textit{0.00060} & \textit{0.1725} \\
\hline
25 & 0.00059 & 0.0159 & 0.00057 & 0.0163 & 0.00066 & 0.0186 \\
   & \textbf{0.00061} & \textbf{0.0610} & \textbf{0.00051} & \textbf{0.2103} & \textbf{0.00056} & \textbf{0.2964} \\
   & \textit{0.00068} & \textit{0.0502} & \textit{0.00050} & \textit{0.1395} & \textit{0.00082} & \textit{0.1335} \\
\hline
24 & 0.00078 & 0.4100 & 0.00084 & 0.2816 & 0.00079 & 0.2395 \\
   & \textbf{0.00063} & \textbf{0.6047} & \textbf{0.00070} & \textbf{0.4760} & \textbf{0.00075} & \textbf{0.2780} \\
   & \textit{0.00141} & \textit{0.1958} & \textit{0.00081} & \textit{0.1746} & \textit{0.00124} & \textit{0.1795} \\
\hline
23 & 0.00090 & 0.2747 & 0.00089 & 0.3165 & 0.00099 & 0.2324 \\
   & \textbf{0.00085} & \textbf{0.3284} & \textbf{0.00089} & \textbf{0.3519} & \textbf{0.00098} & \textbf{0.2403} \\
   & \textit{0.00128} & \textit{0.2393} & \textit{0.00140} & \textit{0.2578} & \textit{0.00112} & \textit{0.1956} \\
\hline
\end{tabular}
\end{table}

Figure~\ref{fig:trans reconstruction det2 from 2n sim} shows the reconstruction based on only the two simulated transits that correspond to our actual data set for WASP-107 b. We clearly see that artifacts associated with the broad stellar lines are heavily amplified, and that reconstruction of the top two orders, dominated by water absorption, essentially failed. The failure is quantified by the peak of the cross-correlation function (CCF) for this order presented in Table~\ref{fig:trans reconstruction det2 from 2n sim} (0.0178 for two nights versus 0.2985 for seven in case of the middle detector). This is not surprising considering the density and strength of water lines in the telluric spectrum, and difficulty distinguishing between the stellar and telluric spectra using just two transits. The two bottom spectral orders containing CO lines are successfully recovered in both simulations thanks to lower strength of CO lines in the stellar spectrum and fewer, more well-separated telluric lines.

An abundance of small-scale features can be seen in the seven-transit reconstruction (black line in Figure~\ref{fig:trans reconstruction det2 from 7n sim}) compared to the two-night test. The inversion code uses the Tikhonov regularization (specified in Section~\ref{sec:forward_model:degeneracies}) for unknown functions to prevent nonphysical values (e.g. negative values of the blocking area). The impact of regularization is most significant for the transmission spectrum, where it controls the trade-off between the S/N and spectral resolution of the result. The relative weight of the contribution of regularization is higher for the two-transit case, as there are fewer observational data.

The right panel of Figure~\ref{fig:ccf versus number of transits} shows how the quality of the transmission spectrum reconstruction improves with the number of transits. The quality of the solution is estimated through the cross-correlation between the TSD solution and the input transmission. This curve will have a different shape if we construct it using separate orders. For example, it will start higher for the two bottom orders in Figure~\ref{fig:trans reconstruction det2 from 7n sim} dominated by CO, while for the two top orders dominated by water, the rise will not start until the fifth transit is added. We also see that the addition of the seventh transit results in a minor improvement of the solution, giving some hints for potential observing strategies.

To test how strongly the solution depends on the initial conditions, we compared the results of TSD runs with two initializations. In the first, we set the stellar flux to the observations at a randomly selected phase just outside the transit, while telluric absorption was set to zero. For the second we set the
tellurics to the observation at the minimum airmass for each transit, while stellar flux had no spectral features. The results
are presented in the bottom panel of Figure~\ref{fig:trans reconstruction det2 from 7n sim}, which also shows the impact of
regularization. The solutions starting with ``all tellurics'' and ``all stellar'' are similar but not identical, giving another view on the precision of
the reconstruction. Reducing the regularization parameter as expected increases the amplitude of the high-frequency noise in the solution.

\section{TSD Performance on Observed Data}\label{sec:spectra}

The previous section presented numeric experiments that demonstrate the ability of an inverse model like TSD to recover high-resolution transmission spectra from realistic quality data \textit{without} assumptions about the exoplanet atmosphere. In this section, we present an example of TSD's performance on two \textit{real} transits of WASP-107 b as observed with CRIRES+. Note that since this paper is intended primarily to present the method itself, we defer the presentation and interpretation of astrophysical results to follow-up studies. 

\subsection{Observations and Data Reduction}\label{sec:spectra:reductions}
CRIRES---the CRyogenic InfraRed Echelle Spectrograph---is an Adaptive Optics (AO) fed \citep{paufique_macao-crires_2004}, high-resolution (R${\sim}$100\,000 when using a 0.2" slit), long-slit echelle spectrograph operating in the $YJHKLM$ ($0.92-5.3\,\mu$m) bands on ESO's Very Large Telescope (VLT) at Cerro Paranal, Chile. The original version of the instrument was in operation between 2006 \citep{kaeufl_crires_2004} and 2014. After decommissioning, CRIRES underwent a substantial upgrade before returning to service in 2021, gaining cross-dispersion optics, state-of-the-art detectors, a new version of the MACAO AO, and a + to its name denoting its new identity as CRIRES+ \citep{dorn_crires_2023}.  Although CRIRES always inhabited a relatively unique instrumental parameter space, a ten-fold increase in the simultaneous wavelength coverage, better throughput, and better reduction pipeline has made CRIRES+ a much more efficient and competitive instrument.

Each CRIRES+ spectral order is projected over three HAWAII-2RG detectors, with the number of orders depending on the spectral band. This allows CRIRES+ to cover each of the $YJHKLM$ bands in a single exposure, albeit with gaps between adjacent detectors, and not the whole free spectral range for orders in redder bands. Wavelength calibrations for the $YJHK$ bands are achieved through the use of a UNe lamp and a Fabry-P\'erot etalon system, which are part of ESO's daytime calibration routine taken as part of the ESO program 60.A-9051(A).

CRIRES was used for the characterization of exoplanetary atmospheres in transmission or emission even before its upgrade \citep[e.g.][]{birkby_detection_2013, de_kok_detection_2013, brogi_carbon_2014}, with the science continuing post-upgrade given its greater wavelength coverage \citep[e.g.][]{yan_crires_2023, lesjak_retrieval_2023, ramkumar_high-resolution_2023,Grasser_2024_crires,parker_2025_crires}. The molecular species present in exoplanetary atmospheres at NIR (Near Infra-Red) wavelengths---such as H$_2$O, CO$_2$, CO, and CH$_4$ to name a few---tend to present as dense forests of absorption features rather than single discrete strong lines, and may typically be detected by cross-correlation with a spectral template.

In our example, we use data from two transits of WASP-107 b, observed as part of Guaranteed Time Observations (GTO) awarded to the CRIRES+ Consortium. These data were already analysed with the cross-correlation technique by \cite{Linn_WASP107_2025}. The observations were taken using the K2148 wavelength setting, with a series of pre-, in-, and post-transit exposures, and used an ABBA nodding pattern to enable for subtraction of the sky background and detector artifacts. The observations are summarized in Table \ref{tab:obs_tab}.

The data were reduced using the standard ESO CRIRES+ DRS (version 2.3.19), available as part of the EsoRex---ESO Recipe Execution---tool \citep{eso_cpl_development_team_esorex_2015}\footnote{\url{https://www.eso.org/sci/software/cpl/esorex.html}}. Calibrations were reduced using the EsoRex commands \texttt{cr2res\_cal\_dark}, \texttt{cr2res\_cal\_flat}, and \texttt{cr2res\_cal\_wave}. The science frames were reduced using the recipe \texttt{cr2res\_obs\_nodding} for each A/B nodding pair with the following (non-standard) parameters: swath width 800, oversample 10, extraction height 21, slit function smoothing 0.5, no spectrum smoothing or cosmic hit removal. The processing chain applied to the science frames therefore consists of the subtraction of the two nodding positions, optimal extraction of the spectra \citep{piskunov_optimal_2021}, correction for the blaze function derived from the flat field, wavelength calibration that accounts for the curvature of the slit image, and resampling of the A and B spectra onto a common wavelength grid. We stress that \textit{no continuum normalization is applied}: the spectra entering TSD are extracted, blaze-corrected fluxes in instrumental units, and the per-phase scaling $S_j$ and offset $D_j$ of Equation~\ref{eq: forward model} are the only place where the overall level of each exposure is adjusted. The blaze correction is a calibration of the instrumental response measured from the flat field, not a fit to the stellar continuum of the science spectra.

\begin{table}
\centering
\caption{Summary of VLT/CRIRES+ observations for WASP-107 b.}
\begin{tabular}{cccc}
Parameter & Unit & Night 1 & Night 2\\
\hline
Date (UTC) & - & 2022-03-11 & 2023-02-23\\
Time (UTC) & - & 03:54--09:30 & 04:55--09:16\\
$N_{\rm exp}$ (in/out) & - & 64 (34/30) & 50 (34/16)\\
$t_{\rm exp}$ & sec & 300 & 300\\
Airmass & - & $[1.03,1.59]$ & $[1.03,1.21]$\\
SNR & - & 133 & 130\\
Resolution & & 100,000 & 100,100 \\
RV$_{\rm bary}$ & \,km\,s$^{-1}$ & $[-11.55,-10.89]$ & $[-18.84,-18.33]$\\
$\Delta$RV$_{\rm bary}$ & \,km\,s$^{-1}$ & 0.01 & 0.01\\
RV$_{\rm planet}$ & \,km\,s$^{-1}$ & $[-6.12,6.09]$ & $[-6.18,6.07]$ \\
$\Delta$RV$_{\rm planet}$ & \,km\,s$^{-1}$ & 0.37 & 0.36\\
\end{tabular}
\label{tab:obs_tab}
\end{table}

\begin{figure*}[h]
    \centering
    \includegraphics[width=0.49\linewidth, trim=3.1cm 2.2cm 1.2cm 2.2cm]{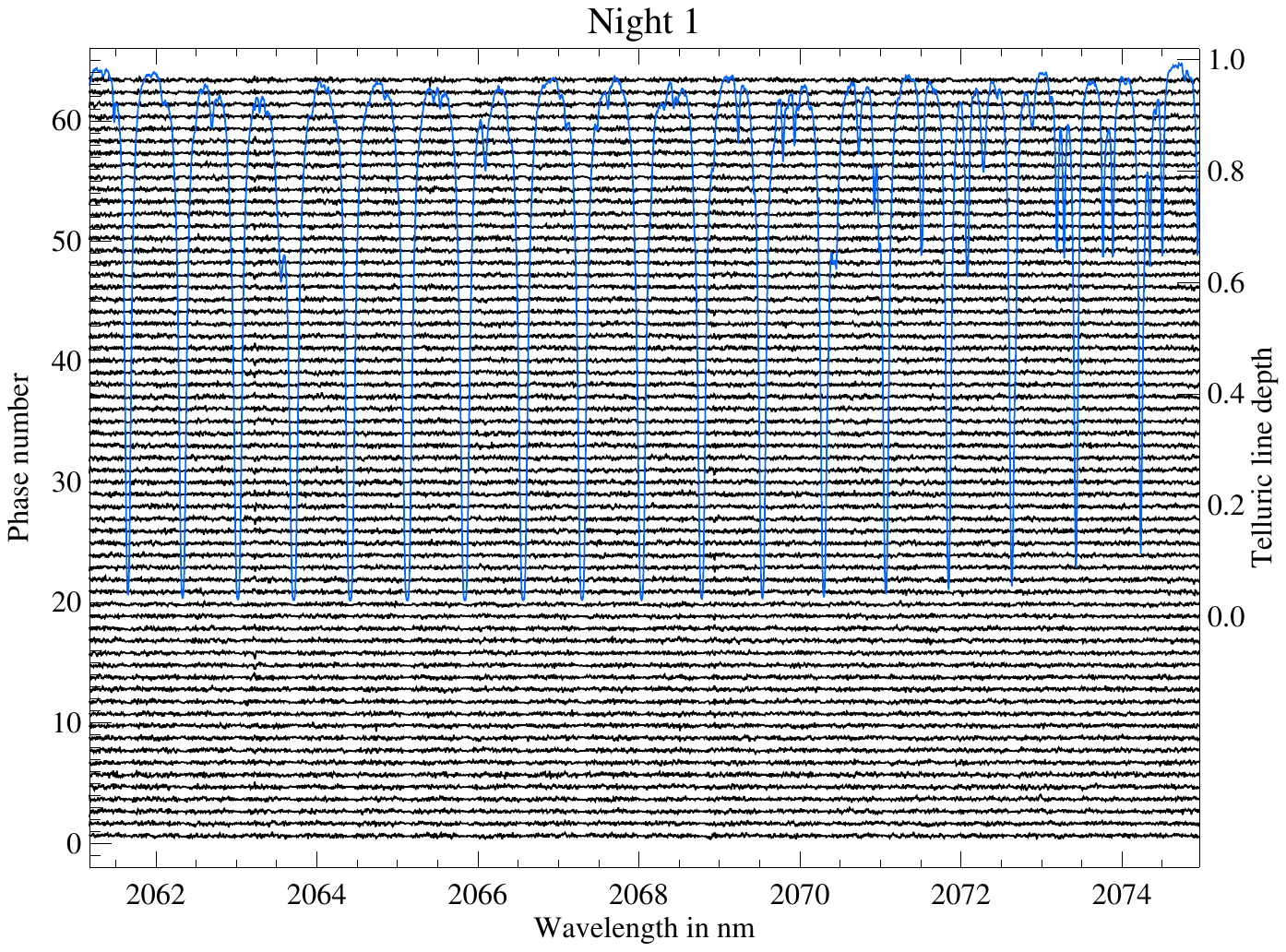}
    \includegraphics[width=0.49\linewidth, trim=3.1cm 2.2cm 1.2cm 2.2cm]{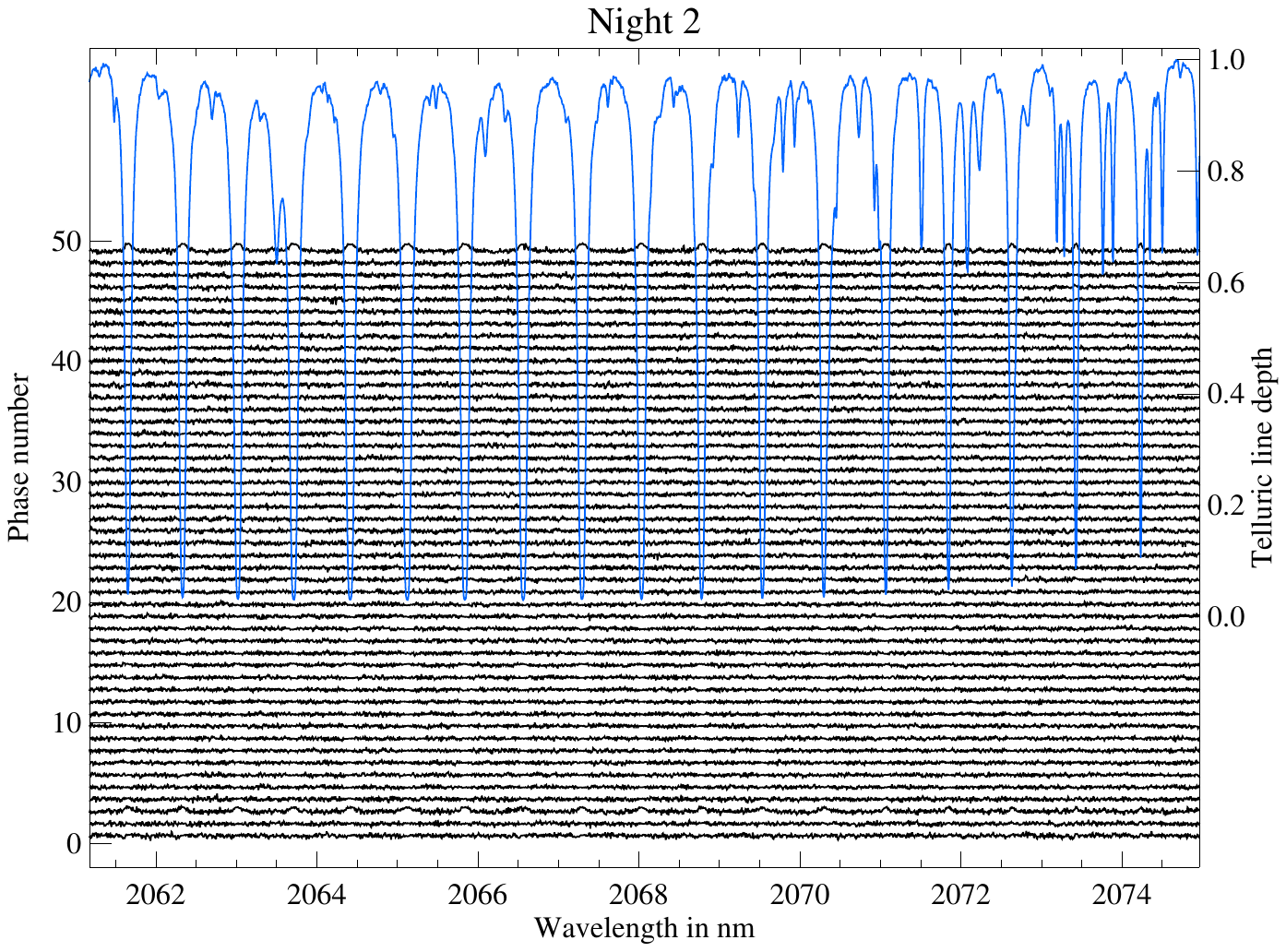}
    \caption{Per-phase differences (black lines) between observed WASP-107 spectra and a TSD-reconstructed telluric absorption with a fixed PSF profile, scaled to appropriate airmass. Vertical offsets between each subsequent phase were added for visibility. A blue overplotted spectrum shows the mean telluric spectrum for each night at the lowest airmass. Note that substantial structure in the residuals is only visible for two phases on night 2 (third from bottom, and topmost), which is indicative of the PSF changes in those phases, but also to the general stability of the PSF otherwise. Both of these phases lie outside the transit.}
    \label{fig:PSF stability}
\end{figure*}

\subsection{Data Preparation}
For use with the inverse method, the reduced data from both nights needed to be cleaned first and formatted into a spectral datacube with a single uniform wavelength scale of shape $[N_{\rm phase}, N_{\rm spec}, N_{\rm px}]$ where $N_{\rm phase}$ is the total number of phases observed across \textit{both} nights; $N_{\rm spec}$ is the number of spectral segments equal to $N_{\rm order} \times N_{\rm detector}$; and $N_{\rm px}$ is the number of spectral pixels per detector. We also compute and store `timestep information' associated with each phase such as the positional and velocity information associated with the system (i.e. barycentric, star, and planet), the fraction of the planet currently in transit, and the projected $\mu$ value for computing limb darkening---all of which are necessary for the functioning and convergence of the inverse problem as described above.

\begin{figure*}[htb!]
    \centering
    \includegraphics[width=\textwidth, trim=1.4cm 2cm 1.8cm 1.3cm]{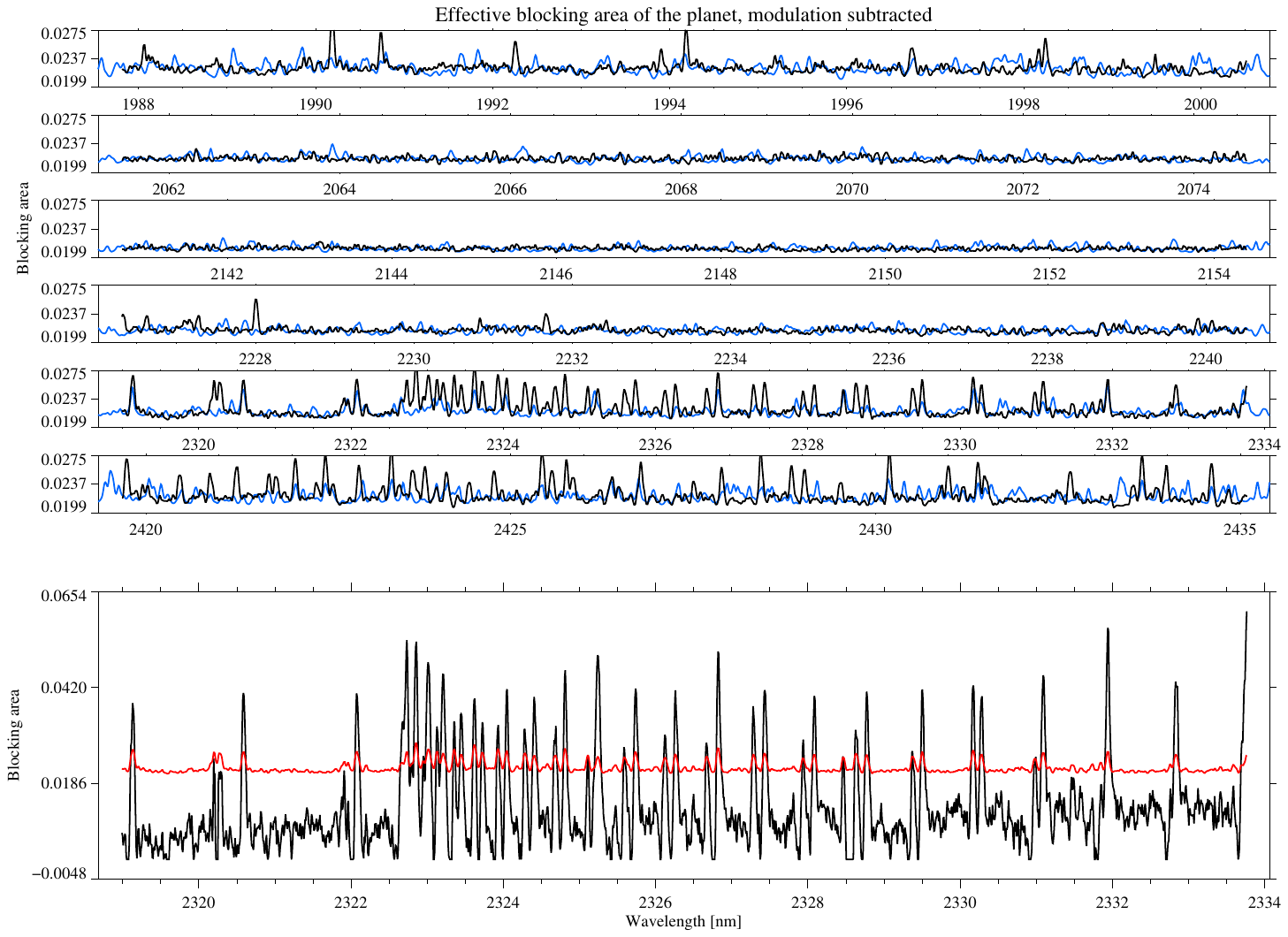}
    \caption{Transmission spectrum of WASP-107 b reconstructed with the TSD method using two transits (black). Only the part of the spectrum from the middle detector is presented for clarity, as in Figure~\ref{fig:trans reconstruction det2 from 7n sim} and Figure~\ref{fig:trans reconstruction det2 from 2n sim}. The petitRADTRANS template described in Subsection \ref{sec:simulator:system_info:planet} (the same as in Figure~\ref{fig:trans reconstruction det2 from 2n sim}) is shown in blue as a reference. Note that since we do not know what the `true' transmission spectrum of WASP-107 b looks like in reality, this template should be taken as a reference only and not the ground truth as in our simulated observations. The six upper panels show the reconstruction after subtraction of the low-frequency modulation discussed in the text, on a common vertical scale and without any rescaling, so that the recovered structure and the template may be compared in amplitude as well as in position. The bottom panel repeats the fifth order before (black) and after (red) subtraction. The vertical range in this panel is different from the rest of the plot.}
    \label{fig:real trans reconstruction det2}
\end{figure*}

\subsection{Potential and real problems with the data}

Some interpolation is unavoidable, both in the preparation of the data and inside the solver, where the unknown functions are shifted between velocity frames. It is worth noting that CRIRES+ samples the resolution element of 3\,\kms\ with 3.5 pixels, so the cubic-spline resampling neither distorts the line profiles appreciably nor introduces significant correlated errors. We verified this directly on the observed spectra: shifting a spectrum by 5\,\kms\ and back again reproduces the original to better than $10^{-4}$ in relative units (excluding the extrapolated pixels at the edge of the segment), rising to $2\times10^{-3}$ in the cores of saturated telluric lines. Both values are well below the photon noise of an individual pixel, which is ${\sim}8\times10^{-3}$. The solver actually avoids extrapolation altogether: only those pixels for which all Doppler-shifted vectors are defined contribute to the normal equations, which is why the ends of the spectral segments rely more heavily on the regularization. Our spectra must be interpolated to a common wavelength scale. The reduction pipeline provides an accurate (to better than 100 m\,s$^{-1}$) wavelength solution for each transit but some (significant) shifts may remain when the AO system over-performs in good seeing conditions. We call this effect ``super-resolution'' \citep[see Appendix 2 in][]{nortmann_crires_2025} and it results in shifts of the stellar image across the slit every time CRIRES+ switches between A and B nodding positions. Such shifts are equivalent to additional RV of up to 700 m\,s$^{-1}$. This happens in good seeing conditions when the central peak of the stellar image becomes narrower than the slit, and the slit viewing camera cannot align stellar image position across the slit for A and B nodding positions. This is seen in the spectra as a systematic radial-velocity difference. Because the shift arises from a displacement of the stellar image and not from the dispersion, it is common to all spectral orders recorded in the same exposure. This is why the correction can be determined once and applied to every spectral segment. We measure it where it is best constrained, in the order richest in strong telluric lines (order 27, second order from the top), and then verify the solution on the remaining orders. The transformation (a second order polynomial in RV) is then constructed for all the phases using one telluric line-dominated order at one particular phase as a reference. The super-resolution correction is done separately for each transit. The reference orders are then used to establish the transformation between the different transits and the original data are interpolated to a common wavelength grid. No sigma clipping is applied to the data cube. The great majority of hot and dead pixels are already eliminated by the subtraction of the nodding pairs and by the optimal extraction, which weights each pixel by the spatial profile and rejects discordant ones \citep{piskunov_optimal_2021}. The few artefacts that survive are typically located at the edges of the spectral segments as a consequence of the slit image curvature. The ESO pipeline reduction flags the affected columns as invalid, and because the projected slit image is tilted the number of flagged columns varies from order to order and between the two ends of a segment. Rather than carrying a different wavelength range in every order, we adopt for each detector the range of columns that is valid in all of its orders. This leaves 2005 wavelength points per spectral segment, the number used throughout this paper for the real data. One further defect deserves mention: a narrow, intermittently noisy region affecting the B nodding position but not A. We replaced the corresponding interval, 32 pixels wide (two readout channels), with the A spectra. Because this repair is applied to alternate exposures only, its imprint is visible in the residuals of orders 4 and 5. The resulting data cube and timestep information table are saved to a FITS file. We note that the data preparation described by \citet{Linn_WASP107_2025} has evolved significantly since that
publication.

Another potential problem is possible changes of the telluric opacity and/or the PSF of CRIRES+ during a single transit. The AO system of CRIRES+ tends to keep the shape of the central peak rather constant as long as the seeing conditions do not change dramatically but it will not correct the changes in opacity due to e.g. changes in humidity. Fortunately, we have ways to detect such changes. First, the reduction pipeline saves the PSF of the spectra for each spectral segment (6 spectral orders split between 3 detectors).
Second, we can detect the changes of the PSF and telluric opacity by subtracting the telluric spectrum derived by TSD for a given transit from the spectral orders dominated by telluric lines, as shown in Figure \ref{fig:PSF stability}. The TSD model does not include any PSF or differential opacity variations and systematic non-flat signatures in the difference spectra would be indicative of such variations.

\subsection{Results from Real WASP-107 b Data}\label{sec:spectra:results}

TSD results on the real data are presented in Figure~\ref{fig:real trans reconstruction det2}. The regularization parameter for the blocking area $P$ was selected to be 10$^{10}$. Since this parameter is referred to the photometric transit depth (Section~\ref{sec:forward_model:degeneracies}), the weight actually entering Equation~\ref{eq: summary P} is $10^{10}(R_\mathrm{p}/R_\star)^2\approx1.9\times10^8$, about a third of the median value of the denominator of that equation. This value prevents the appearance of high-frequency noise and changing it by a factor of 5 shows no noticeable change of $P$. Beyond that the regularization either fails to suppress the noise or begins to damp spectral features that we believe are real. Reducing the parameter by a factor of 10 has a clear impact on the solution in the form of increased small-scale noise seen in the lower panel of Figure~\ref{fig:trans reconstruction det2 from 7n sim}. Although the broadband level of $P$ could be constrained by photometry, we left its recovery to the inversion, and it is compared with the photometric transit depth in Section~\ref{sec:uncertainty}.

The full characterization of the atmosphere of WASP-107 b will be the subject of a follow-up paper. Here our goal is to illustrate TSD's performance on real data, and this is best done by comparing the analysis of our two real observations with the TSD reconstruction of the simulated data for two transits shown in Figure~\ref{fig:trans reconstruction det2 from 2n sim}. One significant difference from the simulations is the low-frequency modulation of the recovered transmission level. It has a typical scale of 20-40 pixels (much broader than the spectral lines). Positive excursions have larger amplitude, comparable to the mean value of $P$. We interpret this modulation as a consequence of inaccurate blaze correction of the individual spectra. We separate it from the spectral structure by fitting a smooth background to each recovered segment and subtracting it, which leaves the variation on the scale of the spectral lines; the results presented in Figures~\ref{fig:real trans reconstruction det2}, \ref{fig:transmission uncertainties} and~\ref{fig:transmission uncertainty distribution} have been treated in this way. The impact of this procedure is illustrated in the bottom panel of Figure~\ref{fig:real trans reconstruction det2}. A numerical assessment of the quality is given as the bottom number in each cell of Table~\ref{tab:sim results} (either RMS or peak value of the CCF). As one can see, the recovered transmission spectra are different but both do reproduce certain (not necessarily the same) features in the synthetic template, shown in blue. This is not surprising as we do not know if the template actually matches the true transmission spectrum at all wavelengths---a current unknown for \textit{all} exoplanets observed at high spectral resolution. There are also similarities between the two reconstructions: the largest deviations from the template are found in the orders with the strongest telluric features and the RMS/CCF pattern follows the simulated two-transit results (all the parameters for the simulations, except for the spectra, were taken from the real observations). From those simulations, it is evident that two transits are not enough to distinguish between the atmospheric lines of WASP-107 b and strong numerous water lines in the Earth's atmosphere (see top panel in Figure~\ref{fig:telluric and flux reconstruction from 7n sim}). Noting this limitation, we do not claim a clear detection of CO or water in the spectra of WASP-107 b.
The comparison is between two reconstructions rather than against a known answer as the reference template is itself an uncertain model.
What the two results have in common is the pattern of where the reconstruction fails, which follows the telluric absorption, and on that basis we expect the results to improve as additional transits are added. We note that the real data contained a few problems not present in the simulated data. This does not prevent the application of TSD after wavelength homogenization (see the previous Subsection). We emphasize that this experiment demonstrates that TSD returns a stable solution on real data; establishing that the recovered spectrum is an accurate representation
of planetary transmission will require more data, injection--recovery and null tests, interpretation of $P$, etc. TSD analysis of WASP-107 b and a few other transiting gas giants observed with CRIRES+ is the subject of a forthcoming paper.

\section{Uncertainty assessment of the results}\label{sec:uncertainty}

\begin{figure*}[htb!]
    \centering
    \includegraphics[width=\textwidth,trim=0cm 3.3cm 0.8cm 10cm]{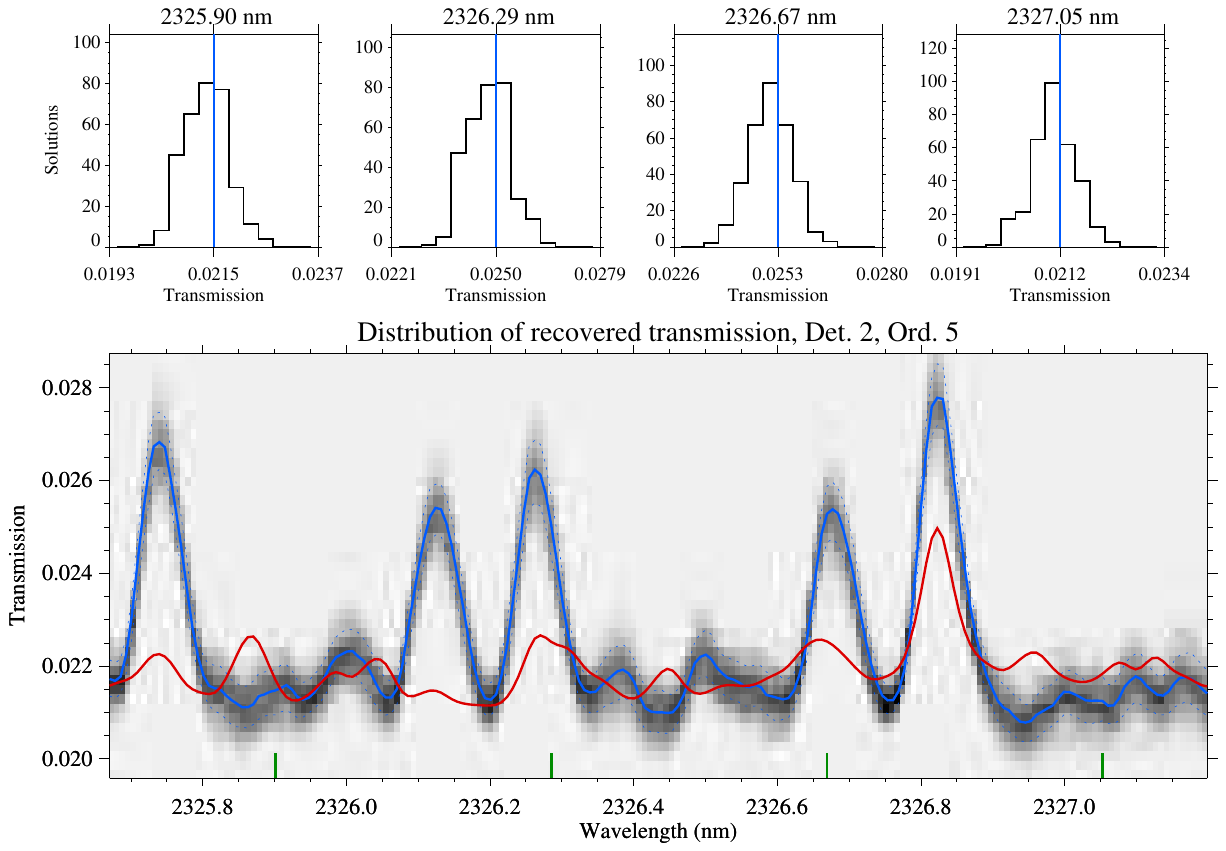}
    \caption{Monte-Carlo distribution of recovered transmission for WASP-107 b, from the ensemble of 320 realizations described in the text, for a region of detector 2 free of strong telluric and stellar lines. \textbf{Bottom:} the grey image shows how many of the realizations return a given transmission at each wavelength, darker being more. The blue solid line is the median of the distribution at each wavelength and the dotted lines enclose its 16th-to-84th percentile range. The red line is a plausible synthetic transmission computed by \citet{Linn_WASP107_2025} for WASP-107 b, the same as shown in Figure~\ref{fig:trans reconstruction det2 from 7n sim}; it is shown for illustration only and is not used in any other way for the reconstruction of $P$. There is no additional scaling, so the two curves may be compared in amplitude as well as in position. \textbf{Top:} the distribution of recovered values at the four wavelengths marked by the short vertical bars at the foot of the lower panel, with the median of the distribution marked in blue.}
    \label{fig:transmission uncertainties}
\end{figure*}

\begin{figure*}[htb!]
    \centering
    \includegraphics[width=\textwidth, trim=0cm 2.cm 0.3cm 7.5cm]{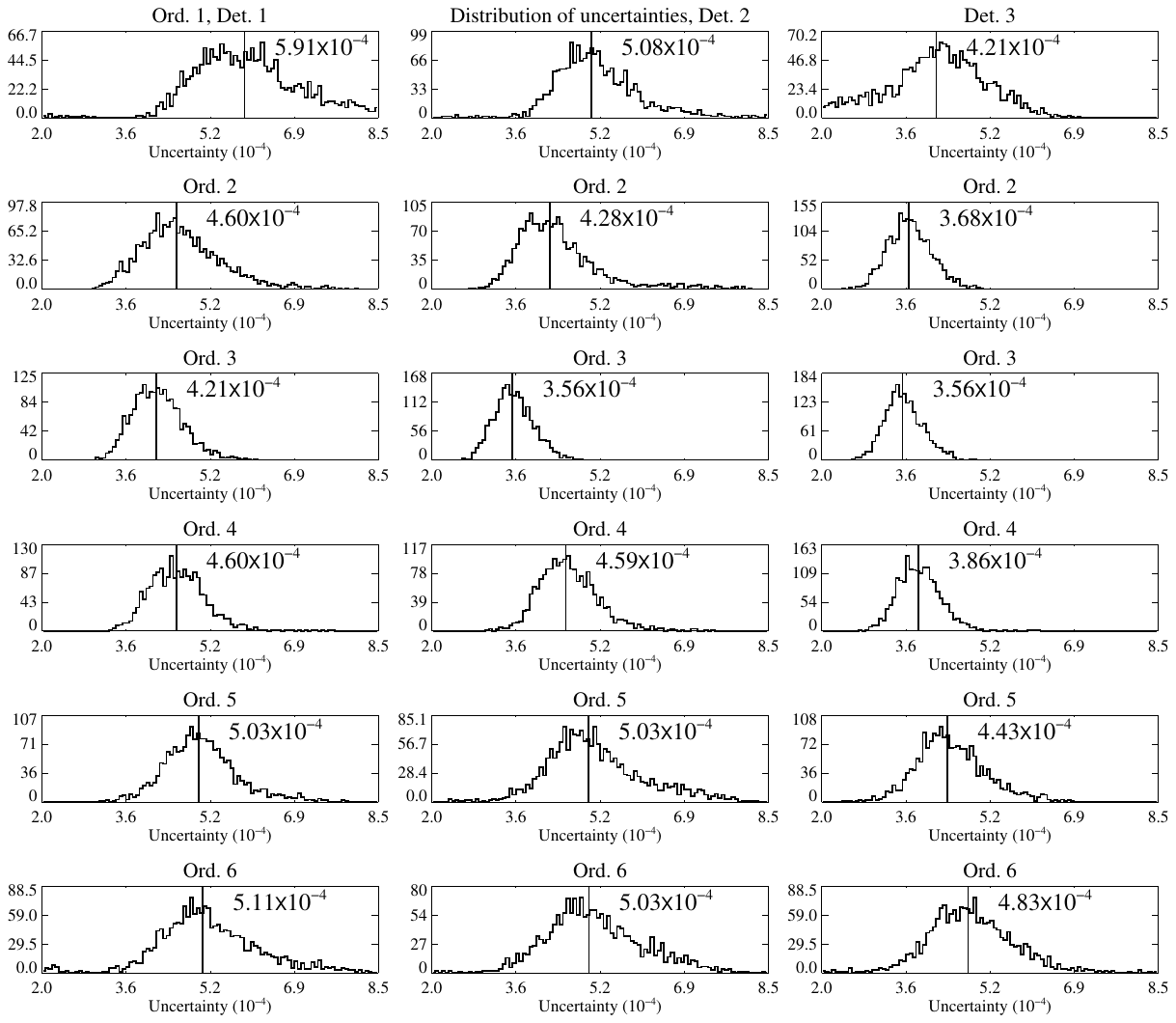}
    \caption{Distribution of uncertainties for the WASP-107 b transmission spectrum reconstructed with TSD. Uncertainties for each spectral point of the transmission vector were derived from the distribution of recovered transmission values at this spectral point. These were combined for all spectral points in a given segment and are shown in corresponding panels as distributions. The quantity shown is the uncertainty of $P_\lambda$ after removal of the slowly varying continuum, which is the part of the reconstruction the smoothing penalty and the data actually constrain; the corresponding figure for the raw $P_\lambda$ is larger by a factor of seven and is quoted in the text. The vertical line and the number in each panel give the median of the distribution, which lies between $3.6$ and $5.9\times10^{-4}$ in units of stellar disk area. All eighteen segments are shown: at the regularization adopted here no segment has more than $4.4$ per cent of its pixels resting against one of the imposed bounds.}
    \label{fig:transmission uncertainty distribution}
\end{figure*}

There is no simple way of propagating the uncertainties through the TSD process analytically. We therefore estimate them by Monte-Carlo propagation of the measurement uncertainties through the whole inversion. The observations have uncertainties established by the 
instrument data reduction pipeline associated with every spectral pixel and every phase. The simplest approach is to perturb the observations drawing from a Gaussian distribution with the width set by these uncertainties. Thus the amplitude of the perturbation of a given spectral pixel in a given phase is set by the uncertainties of the data. We use modified observations to re-run the TSD process and construct the distribution of the recovered values for each element of the unknown functions. Figure~\ref{fig:transmission uncertainties} shows a fragment of the results for the transmission spectrum based on an ensemble of 320 realizations. Each realization perturbs the observations by the pipeline uncertainties and re-converges the full solution starting from the unperturbed one; the perturbations are small, so the solution moves locally rather than having to rediscover the minimum from the initial guess. The broadband level of $P_\lambda$ is not fixed externally in these runs but left to the inversion, so the ensemble measures its uncertainty along with everything else.

Combining TSD runs for perturbed observations allows us to construct the distribution of recovered values for each element of the transmission spectrum ($P_\lambda$). Examples of such distributions are shown in the top panels of Figure~\ref{fig:transmission uncertainties} for four selected wavelengths. The median of each distribution gives the recovered value, shown as the blue line in the bottom panel of Figure~\ref{fig:transmission uncertainties}, and half of its 16th-to-84th percentile range gives the uncertainty. For a Gaussian sample the latter is the standard deviation, and where the distributions are well behaved it agrees with a Gaussian fit to within a few per cent.

At this point, we can construct the probability distribution of the uncertainties for $P_\lambda$ for each spectral segment containing 2005 wavelength elements. This is shown in Figure~\ref{fig:transmission uncertainty distribution}. While the distributions do change, reflecting the level of success of the TSD in each spectral segment, two uncertainties have to be quoted rather than one. As described in Section~\ref{sec:spectra:results}, the recovered $P_\lambda$ of these data carries a slow undulation of much larger amplitude than its line structure, and the two are determined to very different precision. For the raw reconstruction the median per-pixel uncertainty is $3.4\times10^{-3}$ in units of the stellar disk area ($3.5\times10^{-3}$, $3.3\times10^{-3}$ and $3.3\times10^{-3}$ for the three detectors). Once the slowly varying continuum is removed the median falls to $4.5\times10^{-4}$ ($4.8\times10^{-4}$, $4.6\times10^{-4}$ and $4.0\times10^{-4}$), with segment medians between $3.6$ and $5.9\times10^{-4}$; this is the uncertainty of the part of the spectrum that carries the planetary lines, and it is the quantity shown in Figure~\ref{fig:transmission uncertainty distribution}. All eighteen segments are retained: at the regularization adopted here the largest fraction of pixels resting against one of the imposed bounds, $P=0$ or $P$ equal to five times the planet-to-star area ratio, is $4.4$ per cent. These per-pixel statistical uncertainties are what the Monte-Carlo experiment delivers alongside the reconstruction. We deliberately refrain from converting them into a detection significance for the recovered spectrum. A significance attached to an individual spectral feature would require, in addition, the amplitude of the feature and the number of correlated pixels over which it extends, and both belong to the interpretation of the spectrum, which is the subject of the forthcoming paper dedicated to the real observations. For orientation only, the line structure surviving the continuum removal has an amplitude of a few per cent of the blocking area itself, reaching $15$ per cent in the noisiest segment.

The broadband level of $P_\lambda$ deserves separate comment, because it is recovered here rather than imposed. Taken over the eighteen segments its median is $2.12\times10^{-2}$, or $0.97$ of the photometric transit depth, and fourteen of the eighteen lie within a factor of two of that depth. This is a useful check that the inversion is not merely reproducing an externally supplied number, and that the CRIRES$^+$ observations and their reduction preserve this information to a large extent. It is not, however, a competitive measurement of the transit depth: the level of an individual segment has a Monte-Carlo uncertainty of only $4\times10^{-4}$, about two per cent, while the segment-to-segment scatter of the recovered levels is $1.6\times10^{-2}$, some forty times larger. Each segment therefore determines its own level precisely and the segments disagree with one another, which is the signature of a systematic rather than of noise, possibly related to the slow undulation described in Section~\ref{sec:spectra:results}. Photometry remains the appropriate source of the absolute level; what the inversion supplies, and what this paper reports, is the spectral structure.

The assumption of Gaussian distributions throughout the analysis presented in this section does not incorporate the systematic problems present in real-life observations. These include instrumental defects (infrared detectors in particular), imperfect operation of the AO system responsible for keeping constant PSF, relative variations of telluric opacity components, etc. Some of those can be included in the TSD model and we are working on this (e.g. variable telluric opacities), and some may be included in the Monte-Carlo analysis. We also note that the perturbations are drawn from the pipeline pixel uncertainties alone, so the resulting distributions measure the propagation of the photon and detector noise of the observations and nothing else. In particular they do not sample the error of the stellar model, which is instead bounded in Section~\ref{sec:forward_model:specific intensity} and propagated end-to-end by the simulations of Section~\ref{sec:simulator}. These questions will be revisited in an upcoming paper dedicated to real data applications.

\section{Discussion and Conclusions}\label{sec:discussion}

We have presented a novel method for recovering the high-resolution transmission spectrum (effective blocking area of a planet) from high-resolution transmission spectroscopy observations. The new method is formulated as an inverse problem and it reconstructs three main functions: stellar spectrum, telluric spectrum, and the planetary transmission spectrum. The first two can be verified independently against the observations outside transit or observations of a telluric standard, where they show good agreement with expectations. The main benefit of using TSD is that it does not include any model assumptions about the exoplanetary atmosphere, its chemical composition, pressure-temperature gradients, or vertical stratification. This enables model-based analyses to be done on the resulting transmission spectrum, not unlike classical stellar spectroscopy. TSD can naturally combine multiple transits as it in fact requires more than one transit to work. Several transits with well separated barycentric velocities allow planetary features to be recovered even if they overlap with strong telluric lines. In our simulations of WASP-107 b five transits were required before the reconstruction of the transmission spectrum stabilised. Of course this number will depend on the transit depth, telluric strength and pattern, and on the spread of barycentric velocities. High-resolution transmission spectra recovered with TSD offer the possibility to measure chemical composition but also atmospheric dynamics in the presence of globally coherent supersonic winds. The analysis will require a planetary model and a radiative transfer code capable of accounting for complex stratification and 3D dynamics, such as NASA PSG \citep{2018JQSRT.217...86V}.

\subsection{TSD Limitations}\label{sec:discussion:limitations}

TSD has a number of limitations---some are due to the current implementation, and some are implicit to the method. Here is a non-exhaustive list with possible corrective actions:
\begin{itemize}
    \item The assumption of constant transmission spectrum for all the phases may not work for some tidally-locked planets. We are considering introducing day- and night-side transmission spectra that will be combined in the model according to the viewing geometry at each phase.
    \item The relation between stellar flux and specific intensities at a given limb angle $\mu$ was derived assuming a non-rotating star, which excludes the Rossiter--McLaughlin effect from our forward model. The assumption can be relaxed by replacing the analytical integration in Equation~\ref{eq: flux and specific intensity} with numerical integration that incorporates surface motion such as rotation, a simple radial-tangential macroturbulence, and even differential rotation.
    \item The assumption of constant telluric spectrum throughout a single transit may not be realistic, e.g. when humidity changes significantly. As shown by Figure \ref{fig:PSF stability}, we have been lucky with our data in this regard. In the case of less favourable conditions, one possible solution is to split the telluric opacity vector into water and ``the rest'', allowing the water part to change smoothly with time in the same way for all wavelengths.
    \item The uncertainties of the unknown functions were tentatively studied in this work, and even though only the transmission results were shown, the same procedure provides the uncertainties for all the unknowns. The upcoming Python version of the TSD code will have a Monte Carlo post-processing part dedicated to deriving robust uncertainties.
    \item The broadband level of the transmission spectrum is not sufficiently constrained by the spectroscopy itself: it is sensitive to the normalization of the stellar flux, the slit losses, and data reduction. We can fix the mean level using the photometric transit depth but the slow high-amplitude undulations remain. We consider this to be the largest challenge for data preparation and we will invest time and effort in solving it.
    \item The amplitude of the spectral features in the recovered
transmission spectrum does not match the petitRADTRANS template we compare with.
In the raw solution the discrepancy is an order of magnitude, but most of it is
the slow modulation discussed in Section~\ref{sec:spectra:results}: once that is subtracted
the recovered features stand between $0.7$ and $2.0$ times the amplitude of the
template. We caution that this residual ratio is not a
quantity one can quote independently of the treatment applied: filtering the
template in the same way as the reconstruction removes most of its amplitude,
because the template is smooth where the reconstruction carries noise, and the
ratio then takes an altogether different value.
Two explanations remain for the part that is left. The comparison template is based on a planetary atmosphere model, and
it carries all the associated deficiencies and degeneracies:
\citet{2025A&A...701A.133K} examine the reliability of one-dimensional
radiative-convective photochemical-equilibrium retrievals applied to transit
spectra of WASP-107\,b specifically, and find that the inferred atmospheric
properties are subject to substantial systematic and degenerate uncertainties.
A disagreement in amplitude with such a template is therefore not by itself
evidence of an error in the recovery. Such difficulties are not peculiar to
our analysis as shown by \citet{Linn_WASP107_2025}, who apply
cross-correlation of the same observations of WASP-107 b with the same template
as we use for the simulations of Section~\ref{sec:simulator} and achieve only marginal
detections of some molecular species.
Alternatively, TSD may be trying to use $P$ to compensate for inconsistency of the observations with the model.
    \item The most fundamental limitation of TSD is the need to resolve the change of exoplanet RV during a transit. This is typically not a problem for hot Jupiters, where such changes are measured in \kms; but for Earth analogues, we are looking at a few tens of meters per second. For systems with terrestrial planets orbiting an M dwarf (e.g. TRAPPIST-1), the curvature of the orbit increases (smaller orbit) but the stellar mass decreases, leading to a lower RV gradient in the middle of the transit. As such, more methodological work is required to apply TSD-like methods to characterization of Earth analogues.
\end{itemize}

\subsection{Putting TSD in Context with Detrending Methods}

Before we conclude, it is worthwhile to put an inverse method like TSD in context with the detrending
algorithms that remain the field's method of choice for NIR high-resolution transmission spectroscopy. The process
includes two well-separated stages. The first removes the stellar and
telluric signals from the array of spectra of shape $[N_{\mathrm{phase}}, N_\lambda]$, without any planetary
model, by identifying and subtracting the dominant time-correlated components: principal component analysis,
or the iterative error-weighted scheme of \citet{tamuz_correcting_2005} known as SYSREM, which is related to
principal component analysis but distinct from it in that each datum is weighted by its own uncertainty.
Neither performs any cross-correlation. The second stage is effectively the analysis. It infers a planetary signal
from the residuals, and it is here that a model of the planetary spectrum enters, either through cross-correlation
with a template or, in the now widely used likelihood formulations, through comparison with a model subjected to the same
detrending operation \citep{Brogi_retrieving_2019, 2020MNRAS.493.2215G, 2022MNRAS.512.4618G,
maguire_high_2024, blain_formally_2024}. That second stage combines the residuals along the \textit{wavelength}
axis, preserving radial-velocity and time information but not the individual spectral features, whereas TSD
assumes the physical model of Equation~\ref{eq: forward model} and combines the data along the \textit{phase}
axis, preserving the spectral information at the price of assuming the unknown functions constant in time.

Several properties favour the detrending route. It works on a single transit, with no need to wait for
further ones. When several are available they are combined after the inference stage, in phase--velocity rather
than phase--wavelength space, so data from different spectral regions or different instruments can be brought
together---something TSD, which solves for functions on a common wavelength grid, cannot do. The methods are
conceptually simple and have been tested over more than fifteen years, and where the planetary template rests
on extensive previous study its assumptions are likely close to reality, particularly for high signal-to-noise
targets such as hot Jupiters with strong atomic or molecular features.

Two limitations bear on the comparison. The first is that the planetary signal is not in fact preserved
by the detrending: the assumption that the planet's large phase-to-phase Doppler shift protects it while the
quasi-static components are removed holds only approximately, and the algorithms attenuate and distort the
planetary spectrum as well \citep[e.g.][]{2019MNRAS.482.4422C, Cheverall_robustness_2023}. This is understood
and treated explicitly in current practice, by injection--recovery calibration or by applying the same operator
to the model before comparison \citep{2022MNRAS.512.4618G, blain_formally_2024}, so it is a quantified effect
rather than an uncontrolled one; it does become more demanding as the field moves towards less massive planets
on longer orbits around fainter hosts, where the signal is weaker and the planetary velocity change across a
transit smaller. The regime in which such systems remain accessible has been mapped by
\citet{2024AJ....167..272C}, and our earlier statement about low-velocity planets was more absolute than the
evidence supports. The second limitation is that no generally agreed objective criterion exists for the number
of components to remove; the choice is usually made by injection--recovery and remains a judgement, which
complicates comparison between analyses.

The regularization strength in TSD is a tunable choice trading spectral resolution against noise. It is
not dissimilar to conventional spectroscopy of faint targets. In some sense it is comparable to selecting the
number of removed components in detrending but they are not equivalent. They fail in opposite directions:
regularization penalises structure, so an over-regularized solution loses real features and an under-regularized
one is noisy, but neither can manufacture a spectral feature the data do not contain, whereas a planetary
template acts in the opposite sense, since the comparison responds wherever the data resemble the template,
so an incorrect template can produce an apparent detection whose significance depends on a model of the
quantity being measured. One choice can only suppress signal; the other can create it. This problem is well
known---the injection and model-reprocessing work cited above exists precisely to control it. Moreover,
increasing the number of transits reduces the role of regularization but it does not affect the quality of detrending
as it is applied to each transit separately. 

The two approaches also do not produce the same kind of result. Cross-correlation returns a
detection significance and a velocity: evidence that a species is present, and at what shift. Given adequate
data, TSD returns the high-resolution transmission spectrum itself as a function of wavelength, with an
uncertainty per resolution element, and the second is not obtainable from the first. A template-free spectrum
is of most value precisely where the reliability of the template is in question, and recent work on this planet
shows that this is not hypothetical: \citet{Linn_WASP107_2025} analyse the same observations against a template
of the same family as the one entering our simulations, and \citet{2025A&A...701A.133K} examine the modelling
behind such templates for WASP-107 b and find a number of problems and degeneracies affecting the inferred
composition. It would therefore be of interest to compare the two approaches self-consistently, in a simulated
framework where the answer is known and on real observations. That comparison is beyond the scope of the
present paper and we intend to carry it out in forthcoming work.

\subsection{Future Prospects}

Several extensions of the method follow directly from the limitations listed above. The most important are the inclusion of stellar rotation, and with it the Rossiter--McLaughlin effect; a two-component telluric model, with water treated separately from the remaining species, which would relax the assumption of a single airmass-scaled optical depth; and separate transmission spectra for the day and night sides of the planet. We also intend to use our simulation package to quantify the effect of active regions along the transit chord on the robustness of the TSD solution, and to establish the limiting planetary mass accessible to the method in anticipation of ELT/ANDES observations. The implementation is currently being ported to Python. Applications of TSD to the CRIRES+ GTO data and to archival observations of other planets will be reported separately.

     
\section*{Acknowledgements}
The authors are thankful to the Knut and Alice Wallenberg Foundation for generous support of this project (KAW Scholarship of NP). A.D.R acknowledges support from  ANID/Fondo ALMA 2024/N°31240064. We also thank Bengt Edvardsson for help with computing the MARCS model and specific intensity spectra for WASP-107, and Thomas Marquart and Alexis Lavail for working with us on optimization of pipeline reduction of CRIRES+ observations.
Based on observations collected at the European Organisation for Astronomical Research in the Southern Hemisphere under ESO programmes 108.C-0267(D), 110.C-4127(D), and 0100.C-0750(D). Finally, we want to mention the whole CRIRES+ Consortium that made this project possible. We would like to thank the anonymous referee for their suggestions which improved the quality of the manuscript.

In this project we used the following software: \texttt{Astropy} \citep{astropy_collaboration_astropy:_2013}, \texttt{iPython} \citep{perez_ipython:_2007}, \texttt{Matplotlib} \citep{hunter_matplotlib:_2007}, \texttt{NumPy} \citep{harris_array_2020}, \texttt{Pandas} \citep{mckinney_data_2010, reback_pandas-devpandas_2020}, \texttt{SciPy} \citep{jones_scipy:_2016, virtanen_scipy_2020}, and \texttt{PyAstronomy} \citep{pya}.


\appendix
\renewcommand{\theHequation}{app.\arabic{equation}}

\section{Linearization of the forward model}\label{app:linearization}

As for any least-squares problem, to find the solution we take partial derivatives of $\Phi$ over free parameters and then set it to 0. However, we have to be careful with the Doppler shifts. For the partial derivative over the wavelength-dependent function $F^\star$, $P$, and $\tau$, we need to shift the model into the restframe of the function we are solving for in order to get the derivative vector matching specific wavelength pixel. This restframe transformation is essential: without it the derivative vectors are misaligned with the wavelength grid of the unknown function and the iteration fails to converge.

The derivatives are straightforward for some of the unknown functions and more involved for others; we write them out in full below for completeness and reproducibility. We begin by re-writing the expression for $\Phi$ in the instrumental restframe:
\begin{eqnarray}
M_{\lambda}^j &=& \left[F_{\lambda\cdot(1-\gamma_j)}^\star
 - I_{\lambda\cdot(1-\beta_j)}^j\cdot P_\lambda\right]\cdot
 e^{-\tau_\lambda\cdot z_j}\cdot S_j + D_j
\label{eq: observation model} \\
\Phi &=& \sum_{\lambda,j} \left[M_{\lambda}^j - O_{\lambda}^j\right]^2.
\label{eq: minimization in the instrument restframe}
\end{eqnarray}

Now we can perform linearization for the scaling and offset parameters that are meant to describe variable seeing conditions. Partial derivatives of $\Phi$ over $S_j$ and $D_j$ can be written explicitly for each phase as they are not affected by the radial velocities:
\begin{eqnarray}
\frac{1}{2}\frac{\partial \Phi}{\partial S_j} &=& \sum_\lambda \left[M_{\lambda}^j
- O_\lambda^j\right]\cdot\left[F_{\lambda\cdot(1-\gamma_j)}^\star
 - I_{\lambda\cdot(1-\beta_j)}^j\cdot P_{\lambda\cdot(1-\delta_j)}\right]\cdot e^{-\tau_\lambda\cdot z_j} \\
\frac{1}{2}\frac{\partial \Phi}{\partial D_j} &=& \sum_\lambda \left[M_{\lambda}^j
- O_\lambda^j\right].
 \label{eq: partial derivatives for S and D}
\end{eqnarray}

For the minimum of $\Phi$, these derivatives should be equal to zero, which allows us to write a system of two linear equations for each pair of $S_j$ and $D_j$:
\begin{eqnarray}
\label{eq: solution for S and D}
S_j\cdot& \sum_\lambda &\left[F_{\lambda\cdot(1-\gamma_j)}^\star - I_{\lambda\cdot(1-\beta_j)}^j\cdot
P_{\lambda\cdot(1-\delta_j)}\right]^2\cdot e^{-2\tau_\lambda\cdot z_j} + D_j\cdot\sum_\lambda \left[F_{\lambda\cdot(1-\gamma_j)}^\star - I_{\lambda\cdot(1-\beta_j)}^j\cdot
P_{\lambda\cdot(1-\delta_j)}\right]\cdot e^{-\tau_\lambda\cdot z_j} = \nonumber\\
=& \sum_\lambda & O_\lambda^j\left[F_{\lambda\cdot(1-\gamma_j)}^\star - I_{\lambda\cdot(1-\beta_j)}^j\cdot
P_{\lambda\cdot(1-\delta_j)}\right]\cdot e^{-\tau_\lambda\cdot z_j} \\
S_j\cdot& \sum_\lambda &\left[F_{\lambda\cdot(1-\gamma_j)}^\star - I_{\lambda\cdot(1-\beta_j)}^j\cdot
P_{\lambda\cdot(1-\delta_j)}\right]\cdot e^{-\tau_\lambda\cdot z_j} + 
D_j\cdot\sum_\lambda 1 = \sum_\lambda O_\lambda^j. \nonumber
\end{eqnarray}
Different orbital phases are only connected in Equation \ref{eq: solution for S and D} through the assumption of constant stellar flux, telluric absorption, and the planetary blocking area, and thus $S$ and $D$ can be separately evaluated for each orbital phase $j$. Both $S_j$ and $D_j$ are additional free parameters of the fit and are solved for at every phase; they are therefore retained explicitly in all of the update equations that follow (for $F^\star$, $P$, and $\tau$). \\

We continue with the flux $F_\lambda^\star$. For this, we will use the forward model in Equation \ref{eq: final specific intensity} with our approximation for limb darkening. We shall also shift to the stellar restframe and assume unresolved stellar rotation $\gamma_j=\beta_j$. In practice, it means $v\sin i \le 1.5$\,\kms\ for observations at spectral resolution of 100\,000.
\begin{eqnarray}
M_{\lambda\cdot(1+\gamma_j)}^j &=& \left[F_{\lambda}^\star - F_{\lambda}^\star\cdot\frac{L_j}{N} \cdot P_{\lambda\cdot(1-\delta_j+\gamma_j)}\right]\cdot e^{-\tau_{\lambda\cdot(1+\gamma_j)}\cdot z_j}\cdot S_j+D_j
\label{eq: observation model in stellar restframe} \\
\Phi &=& \sum_{\lambda,j} \left[M_{\lambda\cdot(1+\gamma_j)}^j - O_{\lambda\cdot(1+\gamma_j)}^j\right]^2.
\label{eq: minimization in the stellar restframe}
\end{eqnarray}

Expression for the partial derivatives of $\Phi$ over $F_\lambda^\star$ is given below:
\begin{equation}
\frac{1}{2}\frac{\partial \Phi}{\partial F_\lambda^\star}=
\sum_j \left[M_{\lambda\cdot(1+\gamma_j)}^j - O_{\lambda\cdot(1+\gamma_j)}^j\right]
\cdot e^{-\tau_{\lambda\cdot(1+\gamma_j)}\cdot z_j}\cdot S_j\cdot\left[1-\frac{L_j}{N}\cdot P_{\lambda\cdot(1-\delta_j+\beta_j)}\right],
\end{equation}
\noindent
and setting it to zero allows us to obtain an expression for the stellar flux at wavelength
$\lambda$:
\begin{eqnarray}
F_\lambda^\star\cdot&\sum_j& \left[1-\frac{L_j}{N}\cdot P_{\lambda\cdot(1-\delta_j+\beta_j)}\right]^2
\cdot e^{-2\tau_{\lambda\cdot(1+\gamma_j)}\cdot z_j}\cdot {S_j}^2= \nonumber \\
=&\sum_j& \left[O_{\lambda\cdot(1+\gamma_j)}-D_j\right]\cdot \left[1-\frac{L_j}{N}\cdot P_{\lambda\cdot(1-\delta_j+\beta_j)}\right]
\cdot e^{-\tau_{\lambda\cdot(1+\gamma_j)}\cdot z_j}\cdot S_j.
\label{eq: stellar flux}
\end{eqnarray}

We obtain an explicit expression for $F_\lambda^\star$ in each wavelength pixel but when solving one should be mindful of the cores of strong telluric lines, where the attenuation removes essentially all of the stellar signal, resulting in degenerate equations of the type $F_\lambda\cdot 0 = 0$. We deal with this by imposing Tikhonov regularization \citep{Tikhonov1995} on the flux vector in Equation~\ref{eq: stellar flux}.\\

The next step is the solution for the planet blocking area $P$. We shift our discrepancy function $\Phi$ to the restframe of the planet:
\begin{equation}
  \Phi = \sum_{\lambda,j} \left[M_{\lambda\cdot(1+\delta_j)}^j - O_{\lambda\cdot(1+\delta_j)}^j\right]^2,
\label{eq: minimization in the planet restframe}
\end{equation}
and take the derivative over $P$:
\begin{equation}
\frac{1}{2}\frac{\partial \Phi}{\partial P_\lambda} = 
-\sum_{j=j_\mathrm{start}}^{j_\mathrm{end}} \left[M_{\lambda\cdot(1+\delta_j)}^j
- O_{\lambda\cdot(1+\delta_j)}^j\right] \cdot I_{\lambda(1-\beta_j+\delta_j)}^j
\cdot e^{-\tau_{\lambda\cdot(1+\delta_j)}\cdot z_j}
\cdot S_j =0. \nonumber
\end{equation}

Note that summation over phases ($j$) is restricted to the transit phases from $j_\mathrm{start}$ to
$j_\mathrm{end}$ as the blocking area $P$ is part of our model only in these phases. As in the solution for the stellar flux, substituting the full forward model retains the additive offset $D_j$, which appears on the observation side of the equation. The expression for $P$ is:
\begin{eqnarray}
&&\sum_{j=j_\mathrm{start}}^{j_\mathrm{end}}  \left[F_{\lambda\cdot(1-\gamma_j+\delta_j)}^\star - I_{\lambda(1-\beta_j+\delta_j)}^j\cdot P_\lambda\right]\cdot 
e^{-2\tau_{\lambda\cdot(1+\delta_j)}\cdot z_j}\cdot {S_j}^2\cdot I_{\lambda(1-\beta_j+\delta_j)}^j = \nonumber \\
&&= \sum_{j=j_\mathrm{start}}^{j_\mathrm{end}} \left[O_{\lambda\cdot(1+\delta_j)}^j - D_j\right] \cdot I_{\lambda(1-\beta_j+\delta_j)}^j\cdot e^{-\tau_{\lambda\cdot(1+\delta_j)}\cdot z_j}\cdot S_j
\label{eq: initial expression for P}
\end{eqnarray}
and after some algebra:
\begin{eqnarray}
P_\lambda=\frac{\sum_{j=j_\mathrm{start}}^{j_\mathrm{end}}
\left[F_{\lambda(1-\gamma_j+\delta_j)}^\star e^{-\tau_{\lambda\cdot(1+\delta_j)}\cdot z_j}\cdot
S_j - O_{\lambda(1+\delta_j)}^j + D_j\right]
I_{\lambda(1-\beta_j+\delta_j)}^j\cdot e^{-\tau_{\lambda\cdot(1+\delta_j)}\cdot z_j}\cdot S_j}
{\sum_{j=j_\mathrm{start}}^{j_\mathrm{end}}  \left[I_{\lambda\cdot(1-\beta_j+\delta_j)}^j\right]^2\cdot 
e^{-2\tau_{\lambda\cdot(1+\delta_j)}\cdot z_j}\cdot {S_j}^2}\label{eq:equations for planet blocking area}
\end{eqnarray}

The above system of linear equations has a diagonal matrix, so we can write the explicit expression for $P$ at each wavelength pixel. This has a danger that the denominator of Equation \ref{eq:equations for planet blocking area} may be zero in the cores of strongest telluric lines, similar to the case of the stellar flux. As for the stellar flux, this is again resolved by applying Tikhonov regularization to $P$ to restrict high-frequency oscillations and degenerate equations. The addition of regularization transforms Equation~\ref{eq: initial expression for P} into a system of linear equations with a tri-diagonal matrix. \\

The last unknown function is the optical depth of telluric absorption $\tau$, which fortunately does not require a reference frame shift. Noting that only the offset term $D_j$ in the model does not carry the telluric attenuation, the derivative of the model is $\partial M_\lambda^j/\partial\tau_\lambda = -z_j\cdot\left(M_\lambda^j - D_j\right)$, and the partial derivative of the discrepancy function looks like this:
\begin{equation}
\frac{1}{2}\frac{\partial \Phi}{\partial\tau_\lambda} \equiv -\sum_j \left(M_\lambda^j - O_\lambda^j\right)
\cdot \left(M_\lambda^j - D_j\right)\cdot z_j = 0.
\label{eq: equation for telluric tau}
\end{equation}
This time, we have a system of non-linear equations because of the quadratic term in $M$. We denote the partial derivative of $\Phi$ with $Y$ and use the Newton-Raphson (NR) algorithm to find the solution:
\begin{equation}
Y\equiv \sum_j \left(M_\lambda^j - O_\lambda^j\right)
\cdot \left(M_\lambda^j - D_j\right)\cdot z_j = 0.
\end{equation}
We are looking for a set of $\tau_\lambda$ that will turn $Y$ to zero for every wavelength.
To find these, we need partial derivatives of $Y$ over $\tau_\lambda$:
\begin{equation}
\frac{\partial Y}{\partial \tau_\lambda}=-\sum_j
\left(2M_\lambda^j - O_\lambda^j - D_j\right)\cdot \left(M_\lambda^j - D_j\right)\cdot z_j^2.
\end{equation}
The NR expressions for $\tau_\lambda$ corrections is given by the first order Taylor expansion:
\begin{equation}
 \Delta \tau_\lambda =
\frac{\sum_j \left(M_\lambda^j - O_\lambda^j\right) \cdot \left(M_\lambda^j - D_j\right)\cdot z_j}
{\sum_j \left(2M_\lambda^j - O_\lambda^j - D_j\right)\cdot \left(M_\lambda^j - D_j\right)\cdot z_j^2}.
\end{equation}
The denominator vanishes only in the cores of saturated telluric lines (both $O$ and $M$ become zero). In this case the nominator is
also zero and so the ratio can be set to zero. The iteration is terminated when the maximum correction to $\tau_\lambda$ falls below that tolerance; segments with intermediately strong telluric absorption converge faster, since the optical depth is then better constrained by the airmass variation.

\bibliography{aronson_method}{}
\bibliographystyle{aasjournal}


\end{document}